\documentclass[aps,pre,floats,superscriptaddress,nofootinbib,floatfix,twocolumn,longbibliography,showkeys]{revtex4-1}
\usepackage{amssymb,amsmath,amsfonts,amsthm}
\usepackage{graphicx}
\usepackage{dcolumn}
\usepackage{bm}
\usepackage{comment}
\usepackage{csquotes}
\usepackage{mathtools}
\usepackage{gensymb}
\usepackage{hyperref}
\usepackage{makecell}
\usepackage{changepage}
\usepackage{multirow}
\usepackage[table]{xcolor}
\usepackage[mathscr]{euscript}
\usepackage{ulem}

\hypersetup{
	colorlinks=true,
	linkcolor=blue,
	filecolor=blue,      
	urlcolor=blue,
	citecolor=blue,
}

\begin{document}

\title{Stability of discrete-symmetry flocks:\\Sandwich state, traveling domains, and motility-induced pinning}

\author{Swarnajit Chatterjee$^{\star}$}
\email{swarnajit.chatterjee@cyu.fr}
\affiliation{Center for Biophysics \& Department for Theoretical Physics, Saarland University, 66123 Saarbr{\"u}cken, Germany.}
\affiliation{Laboratoire de Physique Th{\'e}orique et Mod{\'e}lisation, UMR 8089, CY Cergy Paris Universit{\'e}, 95302 Cergy-Pontoise, France.}

\author{Mintu Karmakar$^{\star}$}
\email{mkarmakar094@ucas.ac.cn}
\affiliation{School of Mathematical \& Computational Sciences, Indian Association for the Cultivation of Science, Kolkata -- 700032, India.}
\affiliation{Wenzhou Institute of the University of Chinese Academy of Sciences, Wenzhou, Zhejiang 325011, China.}
\affiliation{School of Physical Sciences, University of Chinese Academy of Sciences, Beijing 100049, China.}
\affiliation{Departament de F\'isica de la Mat\`eria Condensada, Universitat de Barcelona, Mart\'i i Franqu\`es 1, E08028 Barcelona, Spain.}

\author{Matthieu Mangeat}
\email{mangeat@lusi.uni-sb.de}
\affiliation{Center for Biophysics \& Department for Theoretical Physics, Saarland University, 66123 Saarbr{\"u}cken, Germany.}

\author{Heiko Rieger}
\email{heiko.rieger@uni-saarland.de}
\affiliation{Center for Biophysics \& Department for Theoretical Physics, Saarland University, 66123 Saarbr{\"u}cken, Germany.}

\author{Raja Paul}
\email{raja.paul@iacs.res.in}
\affiliation{School of Mathematical \& Computational Sciences, Indian Association for the Cultivation of Science, Kolkata -- 700032, India.}

\begin{abstract} 
Polar flocks in discrete active systems are often assumed to be robust, yet recent studies reveal their fragility under both imposed and spontaneous fluctuations. Here, we revisit the four-state active Potts model (APM) and show that its globally ordered phase is metastable across a broad swath of parameter space. Small counter-propagating droplets disrupt the flocking phase by inducing a persistent {\it sandwich state}, where the droplet-induced opposite-polarity lane remains embedded within the original flock, particularly pronounced at low noise, influenced by spatial anisotropy. In contrast, small transversely propagating droplets, when introduced into the flock, can trigger complete phase reversal due to their alignment orthogonal to the dominant flow and their enhanced persistence. At low diffusion and strong self-propulsion, such transverse droplets also emerge spontaneously, fragmenting the flock into multiple traveling domains and giving rise to a short-range order (SRO) regime. We further identify a motility-induced pinning (MIP) transition in small diffusion and low-temperature regimes when particles of opposite polarity interact, flip their state, hop, and pin an interface. Our comprehensive phase diagrams, encompassing full reversal, sandwich coexistence, stripe bands, SRO, and MIP, delineate how thermal fluctuations, self-propulsion strength, and diffusion govern flock stability in discrete active matter systems.
\end{abstract}

\maketitle
\def\thefootnote{{$\star$}}\footnotetext{These authors have contributed equally to this work}\def\thefootnote{\arabic{footnote}}

\section{Introduction}
Active matter, a natural or synthetic non-equilibrium system, consists of agents that consume energy to move or exert mechanical forces. Over the past three decades, extensive investigations have established active matter as a prominent area of research~\cite{marchetti2013hydrodynamics,de2015introduction,bechinger2016active,shaebani2020computational,bowick2022symmetry,dauchot2024active,te2025metareview}. Collections of active particles in such systems often display intricate dynamics and collective phenomena, including the formation of large, ordered structures known as flocks. Flocking~\cite{toner2024physics}, the coordinated movement of numerous agents, is a hallmark of many biological and physical systems. It represents a quintessential non-equilibrium behavior commonly encountered in natural settings~\cite{ballerini2008interaction, cavagna2014bird,becco2006experimental,calovi2014swarming,steager2008dynamics,gomez2022intermittent}.

The field of active matter physics arguably originated in 1995, when Vicsek and collaborators published their foundational work~\cite{vicsek1995novel} on collective motion, drawing an analogy between flocking behavior and ferromagnetism. Their {\it active XY} model, later referred to as the Vicsek model (VM), displays true long-range order in two dimensions~\cite{toner2012reanalysis,chate2020dry}, seemingly in contradiction with the well-known Mermin-Wagner theorem~\cite{mw1966}. This does not pose a real inconsistency, as the {\it moving} flock represents a far-from-equilibrium system~\cite{ginelli2016physics,toner2024physics}. Later, the active Ising model (AIM)~\cite{solon2013revisiting,solon2015flocking,AIM2023,AIM2024} was introduced, which replaces the continuous rotational symmetry of the VM with a discrete symmetry, although preserves the essential physics of the VM. The AIM exhibits three steady-states similar to the VM~\cite{solon2015flocking}: disordered gas at high noise and low densities, polar liquid at low noise and high densities, and a phase-separated liquid-gas coexistence region at intermediate densities and noise. The key distinction between the VM and the AIM is that the former exhibits giant density fluctuations leading to microphase separation of the coexistence region, while the latter shows normal density fluctuations resulting in bulk phase separation~\cite{solon2015phase}. Recently, inspired by the AIM, other active spin models such as the $q$-state active Potts model (APM)~\cite{chatterjee2020flocking, mangeat2020flocking} and the $q$-state active clock model (ACM)~\cite{chatterjee2022polar,solon2022susceptibility} have emerged as more generalized models of flocking, which can bridge the continuous symmetry VM and the discrete symmetry AIM~\cite{karmakar2024consequence}. Both the APM and ACM are excellent frameworks for investigating collective motion due to their simplicity and capacity to capture key features of flocking behavior, including band-to-lane reorientation transition~\cite{chatterjee2020flocking, mangeat2020flocking}, flocking to jamming transition under volume exclusion~\cite{karmakar2023jamming}, and the role of spatial anisotropy on flocking pattern formation~\cite{chatterjee2022polar,solon2022susceptibility}. All these agent-based models have played an important role in shaping our understanding of polar active matter.

Because of their intrinsically out-of-equilibrium character, active systems are capable of exhibiting long-range order (LRO) and can survive spin wave fluctuations~\cite{gregoire2004onset,bertin2006boltzmann,chate2008modeling,chate2008collective,bertin2009hydrodynamic}. This led to the general belief that polar ordered phases in active matter are typically stable against such fluctuations. However, more recent findings have shown that even minor disturbances, such as a small obstacle, can destabilize a flock, triggering transient dynamics before the system reaches a new steady state~\cite{codina2022small, benvegnen2023meta, karmakar2024consequence}. In a similar vein, significant spontaneous fluctuations have also been observed to disrupt the LRO phase. Notable examples include constant-density Toner-Tu flocks~\cite{besse2022metastability}, one-species~\cite{jd2024MIP} and two-species~\cite{TSAIM} AIM, and the non-reciprocal XY model~\cite{dopierala2025inescapable}, in which the ordered states eventually become metastable and give way to a disordered phase over long periods -- either as irregularly scattered local domains~\cite{jd2024MIP,TSAIM} or as dynamic aster foams separated by shock lines~\cite{besse2022metastability,dopierala2025inescapable}.

Furthermore, the ordered phase of the VM, which breaks a continuous symmetry, has been argued to be metastable in parts of the phase diagram~\cite{codina2022small}, where, below a critical noise level, no dense cluster, regardless of its size, triggers a reversal. In contrast, the ordered phase of the AIM, which breaks a discrete symmetry, exhibits metastability both with artificially nucleated minority-phase droplets~\cite{benvegnen2023meta} and through the spontaneous formation of droplets with opposite polarity~\cite{jd2024MIP,TSAIM}. As a result, the entire flocking phase of the AIM has been proposed to be metastable in the thermodynamic limit~\cite{benvegnen2023meta}. Notably, no spontaneous nucleation or destabilization of the Vicsek liquid phase has been observed so far. This evidence suggests that, unlike in equilibrium systems, flocking models with continuous symmetries may be more robust than their discrete counterparts in non-equilibrium settings~\cite{benvegnen2023meta}. These results raise an important question: Is the stability of two-dimensional orientationally ordered phases in discrete flocking models inherently fragile? If so, under what conditions do these ordered states become stable, and what microscopic mechanisms are responsible for the transition? To explore these questions, it is necessary to investigate a discrete flocking model with more internal degrees of freedom than the AIM. A natural candidate is the more general $q=4$-state APM~\cite{chatterjee2020flocking, mangeat2020flocking}, which includes four spin states in contrast to the two in the AIM.

In this paper, we investigate the stability of the ordered flocking phase in the 4-state APM with artificial droplet excitations~\cite{benvegnen2023meta} and the potential for spontaneous nucleation~\cite{jd2024MIP,TSAIM}. Through detailed numerical simulations, we investigate how different parameters, such as droplet size, droplet density, self-propulsion velocity, diffusion constant, and system size, influence the stability of the polar liquid phase. Based on our findings, we argue that polar flocks in the APM are also metastable across a broad region of the phase diagram, consistent with observations in other discrete flocking models~\cite{benvegnen2023meta, jd2024MIP, TSAIM}.


\section{Model}
\label{ab_model}
We consider an ensemble of $N$ particles defined on a two-dimensional lattice of size $L_x \times L_y$ with periodic boundary conditions. The average particle density of the system is $\rho_0=N/L_xL_y$. The spin state of the $k$-th particle on lattice site $i$ is denoted by $\sigma^k_i = \{1, 2, 3, 4\}$ corresponding to the movement directions right, up, left, and down, respectively, while the number of particles in state $\sigma$ on-site $i$ is denoted $n^\sigma_i$. The local density on-site $i$ is then defined by $\rho_i=\sum_{\sigma=1}^4 n^\sigma_i$. The Hamiltonian of a 4-state APM is defined as $H=\sum_i H_i$ decomposed as the sum of local Hamiltonian $H_i$ ~\cite{chatterjee2020flocking,mangeat2020flocking}:
\begin{equation}
\label{Hapm}
H_i=-\frac{J}{2\rho_i}\sum_{k=1}^{\rho_i}\sum_{l\ne k}\left(4\delta_{\sigma_i^k,\sigma_i^l}-1\right) \, ,
\end{equation}
where $J$ is the coupling between the particles located on the same site. The local magnetization corresponding to state $\sigma$ at site $i$ is defined as 
\begin{equation}
\label{eq:mag}
    m_i^{\sigma}=\frac{1}{3}\left(4n_i^{\sigma}-\rho_i\right) \, .
\end{equation}
A particle at site $i$ with state $\sigma$ can either flip to another state $\sigma^\prime$ or hop to any neighboring sites. Flipping is a purely on-site phenomenon, and from Eq.~\eqref{Hapm}, one can calculate the local energy difference before and after the flipping. The expression of the energy difference reads $\Delta H_i=4J(n_i^{\sigma}-n_i^{\sigma^\prime}-1)/\rho_i$.
The flipping is then accepted with the rate~\cite{chatterjee2020flocking,mangeat2020flocking}
\begin{align}
\label{flipeq}
W_{\rm flip}(\sigma\to\sigma^\prime)&=\gamma \exp\left[-\frac{4\beta J}{\rho_i}\left(n_i^\sigma-n_i^{\sigma^\prime}-1\right)\right] \, ,
\end{align}
where $\beta=T^{-1}$ is the inverse temperature, and without any loss of generality, we consider $J=\gamma=1$.

The biased diffusion mechanism is similar to the process described in Refs.~\cite{chatterjee2020flocking,mangeat2020flocking}. A particle with state $\sigma$ hops to a direction $p$ with rate
\begin{equation}
\label{whop}
W_{\rm hop}(\sigma,p)=D\left[1+\frac{\epsilon}{3}\left(4\delta_{\sigma,p}-1\right)\right] \, ,
\end{equation}
where $D>0$ is the diffusion constant and $\epsilon$ ($0 \leqslant \epsilon \leqslant 3$) is the self-propulsion parameter. At $\epsilon=3$, the particles move purely ballistically, resulting in complete self-propulsion, while $\epsilon=0$ corresponds to the purely diffusive model. The total hopping rate is $4D$.

The simulation evolves in the unit of Monte Carlo steps (MCS) $\Delta t=[4D+\exp(4\beta)]^{-1}$, resulting from a microscopic time $\Delta t/N$, previously used in the simulations of AIM~\cite{solon2015flocking,AIM2024,TSAIM}, APM~\cite{chatterjee2020flocking,mangeat2020flocking}, and ACM~\cite{chatterjee2022polar}. During $\Delta t/N,$ a randomly chosen particle either updates its spin state with probability $p_{\rm flip}=W_{\rm flip}\Delta t$ or hops to one of the neighboring sites with probability $p_{\rm hop}=W_{\rm hop}\Delta t$.  


\section{Results}
\label{result}
The APM~\cite{chatterjee2020flocking,mangeat2020flocking}, proposed as a $q$-state extension of the two-state AIM~\cite{solon2013revisiting,solon2015flocking}, displays a flocking transition from a high-temperature, low-density gaseous phase to a low-temperature, high-density polar liquid phase, passing through a liquid-gas coexistence regime at intermediate densities and temperatures. Within this coexistence region, the fully phase-separated liquid bands exhibit a notable reorientation transition—from transverse motion relative to the average direction of particle movement at low velocities and high temperatures, to longitudinal lane alignment at high velocities and low temperatures. This reorientation is driven by the suppression of transverse diffusion at high velocities, which favors the stability of longitudinal lanes. Such a transition is not observed in the AIM~\cite{solon2013revisiting,solon2015flocking} or the VM~\cite{solon2015phase}, where the bands consistently move in the transverse direction.

In this section, we report the results of a systematic analysis of the stability of the low-temperature, high-density polar liquid phase in the 4-state APM. Our study focuses on two primary scenarios: the response of the APM to droplet excitations (Sec.~\ref{sec:hydro} for the hydrodynamic theory and Sec.~\ref{sec:droplet} for numerical simulations) and the behavior of the APM in the small diffusion limit (Sec.~\ref{sec:sponnucl} only for numerical simulations).


\subsection{Hydrodynamic theory for droplet-induced perturbations}
\label{sec:hydro}

\begin{figure}
    \centering
    \includegraphics[width=\columnwidth]{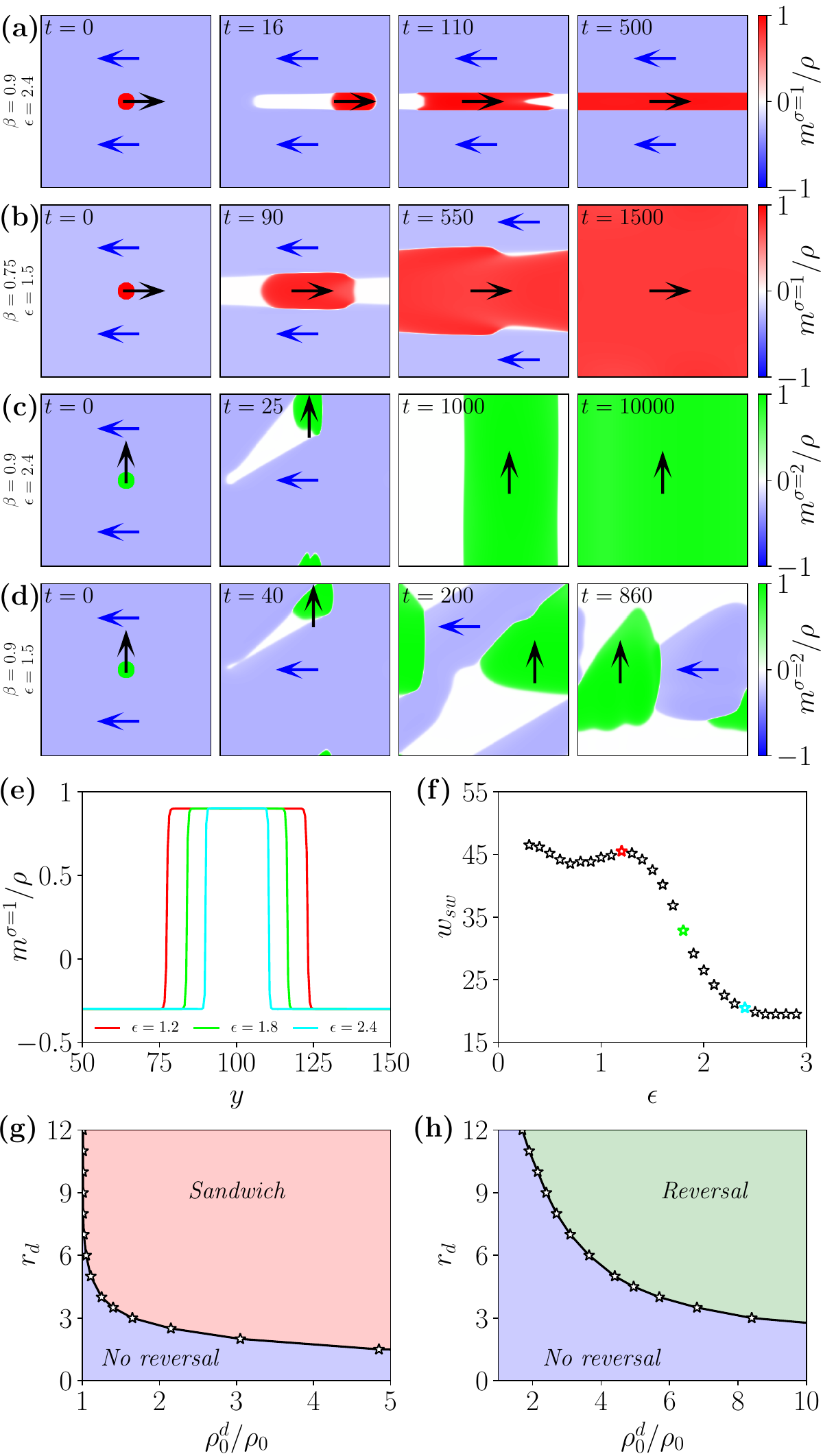}
    \caption{(color online) {\it Hydrodynamic theory of the APM with droplet excitations}. (a--b) Time-evolution snapshots showing (a)~the formation of a sandwich state ($\beta=0.9$, $\epsilon=2.4$), and (b)~the complete reversal ($\beta=0.75$, $\epsilon=1.5$) for a counter-propagating droplet ($\sigma=1$). (c--d) Time-evolution snapshots showing (c)~the complete reversal ($\beta=0.9$, $\epsilon=2.4$), and (d)~the formation of a persistent motion with two orthogonal states ($\beta=0.9$, $\epsilon=1.5$) caused by a transversely-propagating droplet ($\sigma=2$). The initial ordered state is of state $\sigma=3$. Parameters of the droplet: $r_d = 10$ and $\rho_0^d = 5\rho_0$. Arrows indicate the direction of motion, and the colorbar represents the corresponding droplet magnetization. A movie (\texttt{movie01}) of the same can be found at Ref.~\cite{zenodo}. (e)~Magnetization profiles of the sandwich state formed by the counter-propagating droplet ($r_d = 10$ and $\rho_0^d = 5\rho_0$), $\beta=0.9$ and different values of $\epsilon$. (f)~Corresponding sandwich width $w_{\rm sw}$ as a function of $\epsilon$. The red, green, and cyan stars represent the corresponding widths of the profiles shown in (c). (g--h)~$(\rho_0^d,r_d)$ stability diagrams for (g) counter- and (h) transversely-propagating droplets for $\epsilon = 1.2$. Parameters: $D=1$, $\rho_0 = 3$, and $L = 200$.}
    \label{fig:apm_ftcs_snaps}
\end{figure}

We first present our results of the droplet excitation from the refined mean-field coarse-grained hydrodynamic description of the APM presented in Ref.~\cite{mangeat2020flocking}. If we define the average density of particles in the state $\sigma$ at the 2d position {\bf x} as $\rho({\bf x},t) = \langle n_i^\sigma(t) \rangle$, the hydrodynamic equation of the density field $\rho$ for state $\sigma$ is given by~\cite{mangeat2020flocking}:
\begin{equation}
\label{PDEhydro0}
\partial_t \rho_\sigma = D_\parallel \partial_\parallel^2 \rho_\sigma + D_\perp \partial_\perp^2 \rho_\sigma - v \partial_\parallel \rho_\sigma + \sum_{\sigma' \ne \sigma } I_{\sigma \sigma'} ,
\end{equation}
where the interaction term $I_{\sigma \sigma'}$ is
\begin{gather}
I_{\sigma \sigma'} = \left[\frac{4\beta J}{\rho}(\rho_\sigma+\rho_{\sigma'}) -1 - \frac{r}{\rho} - \alpha \frac{(\rho_\sigma-\rho_{\sigma'})^2}{\rho^2}\right](\rho_\sigma-\rho_{\sigma'}). \nonumber
\end{gather}
$D_{\parallel}=D(1+\epsilon/3)$ and $D_{\perp}=D(1-\epsilon/3)$ are the diffusion constants in the parallel, ${\bf e_\parallel} = (\cos \phi,\sin \phi)$, and perpendicular, ${\bf e_\perp} = (\sin \phi, -\cos \phi)$, directions, where $\phi=\pi(\sigma-1)/2$ is the favored angle for state $\sigma$. The self-propulsion velocity $v=4D\epsilon/3$ is along ${\bf e_\parallel}$. Derivatives in these directions are denoted as $\partial_\parallel = {\bf e_\parallel} \cdot \nabla$ and $\partial_\perp = {\bf e_\perp} \cdot \nabla$. In the flipping term $I_{\sigma \sigma'}$, $\alpha= 8(\beta J)^2(1-2\beta J/3)$~\cite{mangeat2020flocking}.

We use explicit Euler forward time centered space (FTCS)~\cite{press2007numerical} differencing scheme to numerically integrate Eqs.~\eqref{PDEhydro0}. We solve the system of four coupled partial differential equations, one for each $\sigma$, on a square domain of size $L \times L$ with periodic boundary conditions applied in both directions. In our simulation, $L=200$ and the maximum simulation time is $t_{\rm sim} = 10^4$. To maintain the numerical stability criteria, we set $\Delta x=0.5$ and $\Delta t=10^{-3}/\Delta x^2$ as the discretization in space and time, respectively. These discretization parameters satisfy the Courant-Friedrichs-Lewy stability condition~\cite{courant1928partiellen}. In our numerical implementation, we fix $D = J = r = 1$, defining a scaling for the time, temperature, and density, and prepare the initial system with a high-density droplet centered and polarized differently from the surrounding homogeneous liquid phase.

Fig.~\ref{fig:apm_ftcs_snaps}(a--d) shows the droplet dynamics for counter- and transversely-propagating droplets at the low temperature polar ordered phase of the APM. At low temperature ($\beta=0.9$) and high velocity ($\epsilon=2.4$), as $D_{\perp} < D_{\parallel}$, the growth of the droplet along the transverse direction is suppressed, which leads to the formation of a partially reversed mixed {\it sandwich state} for a counter-propagating droplet [Fig.~\ref{fig:apm_ftcs_snaps}(a), $t=500$], where regions of the initial ordered phase (blue) coexist with a lane (red) formed by the droplet propagation. However, at higher temperature ($\beta=0.75$) and lower velocity ($\epsilon=1.5$), the counter-propagating droplet grows with a comet-like structure, similar to the one observed for the AIM~\cite{benvegnen2023meta}, and the initial ordered phase is completely reversed [Fig.~\ref{fig:apm_ftcs_snaps}(b), $t=1500$]. Similarly, a complete reversal of the initial liquid phase is observed for a transversely-propagating droplet excitation [Fig.~\ref{fig:apm_ftcs_snaps}(c), $t=10^4$]. Moreover, at large $\epsilon$, the pronounced anisotropy in hopping rates, with $D_\perp \ll D_\parallel$, facilitates the formation of stable longitudinal lanes. However, the expansion of such lanes across the domain is significantly hindered by the suppressed transverse diffusion $D_\perp$, resulting in slow lateral growth. Consequently, the steady state for the droplet of $\sigma = 2$ in panel (c) is reached much later than that for the droplet of $\sigma = 1$ in panel (a). Furthermore, at intermediate velocity ($\epsilon=1.5$), the steady state is composed of two orthogonally moving clusters, with the directions corresponding to the initial ordered state and the inserted droplet. These two clusters never merge into a single flock, even if they frequently collide, due to the square geometry of the periodic domain.

Fig.~\ref{fig:apm_ftcs_snaps}(e) displays the magnetization profiles for a $\sigma=1$ state droplet as a function of the self-propulsion velocity $\epsilon$, obtained by solving the APM hydrodynamic equations at low temperature $(\beta=0.9)$. These profiles are plotted along the $y$-axis by integrating along the $x$-axis. As shown, the widths decrease with $\epsilon$, prompting us to calculate the sandwich state width $w_{\rm sw}$ as $\epsilon$ is varied [Fig.~\ref{fig:apm_ftcs_snaps}(f)].

Fig.~\ref{fig:apm_ftcs_snaps}(g--h) shows the $(\rho_0^d,r_d)$ stability diagrams from the hydrodynamic theory of the APM. We initialize the system in an ordered phase with state $\sigma=3$ and investigate the steady-state behavior as a function of droplet radius ($r_d$) and droplet density ($\rho_0^d$) for $\sigma=1$ counter-propagating droplets [Fig.~\ref{fig:apm_ftcs_snaps}(g)] and $\sigma=2$ transversely-propagating droplets [Fig.~\ref{fig:apm_ftcs_snaps}(h)]. As shown, at large $r_d$ and $\rho_0^d$, the phase space exhibits a sandwich state for counter-propagating droplets, while a reversal of the initial ordered phase is observed for transversely-propagating droplets, similar to the one shown in panel (d).

\begin{figure*}
    \centering
    \includegraphics[width=1.85\columnwidth]{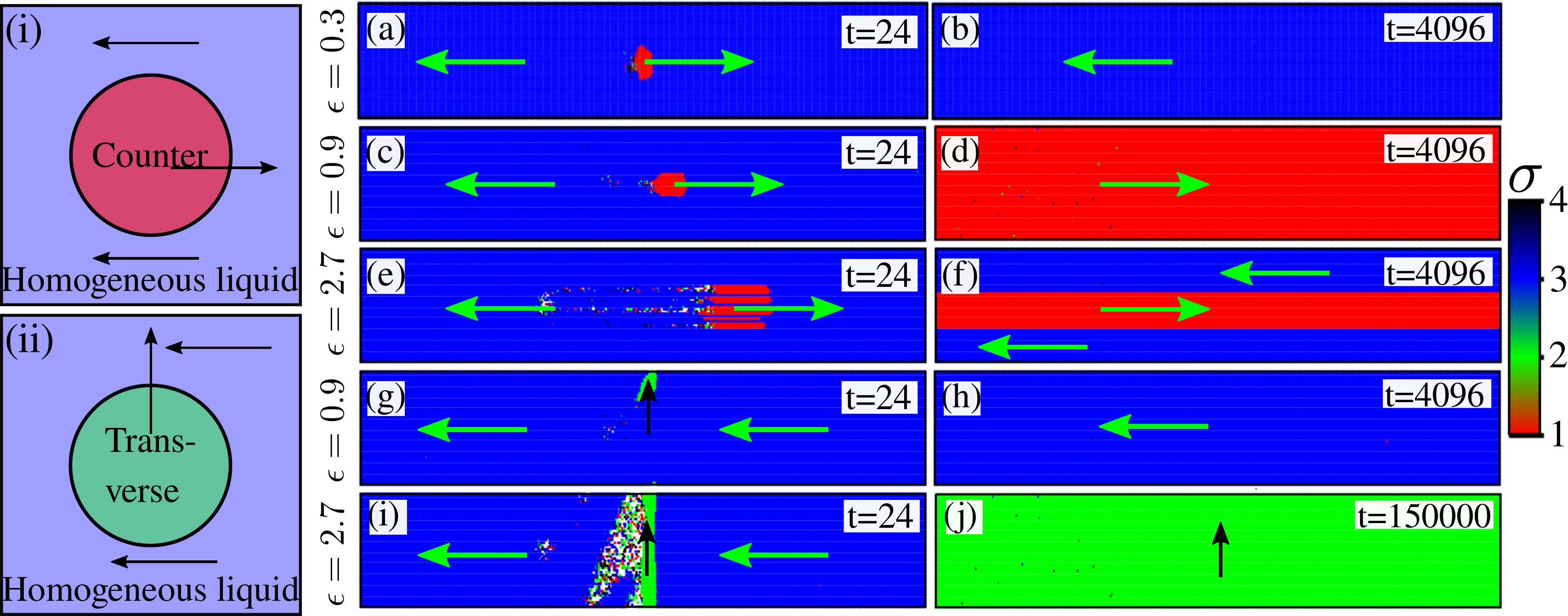}
    \caption{(color online) {\it Time evolution and steady-states with droplet excitations.} (i--ii) Schematic illustrating the movement of droplets: (i) a counter-propagating droplet and (ii) a transversely-propagating droplet, relative to the motion of the homogeneous liquid phase. (a, c, e, g, i) Early-time snapshots and (b, d, f, h, j) steady-state snapshots for (a--f) counter-propagating and (g--j) transversely propagating droplets in the 4-state APM. A counter-propagating droplet with a very small self-propulsion velocity of $\epsilon=0.3$ is unable to reverse the initial liquid phase (a--b) but reverses it for intermediate velocity $\epsilon=0.9$ (c--d). (e--f) Counter-propagating droplet at large self-propulsion velocity $\epsilon=2.7$ creates a sandwich state of $\sigma=3$ (blue) and $\sigma=1$ (red). (g--h) Transversely propagating droplet at small self-propulsion velocity $\epsilon=0.9$ could not reverse the initial liquid phase, although reverses it for large self-propulsion velocity $\epsilon=2.7$ (i--j). Parameters: $D=1$, $\beta = 1$, $\rho_0 = 10$, $r_d = 10$, $\rho_0^d = 1.2\rho_0$, $L_x = 500$, and $L_y = 50$. Colorbar legend: red $(\sigma=1)$: right; green $(\sigma=2)$: up; blue $(\sigma=3)$: left; black $(\sigma=4)$: down. Arrows indicate the direction of motion. A movie (\texttt{movie02}) of the same can be found at Ref.~\cite{zenodo}.}
    \label{fig:fig_drop_fate}
\end{figure*}

\subsection{Numerical simulations with droplet excitation}
\label{sec:droplet}

We now undertake a detailed analysis of the 4-state APM using Monte Carlo simulations, which constitute the core of our study and enable a thorough examination of the stability of the polar-ordered liquid phase. The stability of this phase is assessed on two-dimensional square and rectangular lattices by introducing either a counter-propagating or a transversely propagating liquid droplet into the high-density ordered liquid, in line with the coarse-grained hydrodynamic framework. A schematic diagram of this arrangement is shown in Fig.~\ref{fig:fig_drop_fate}(i--ii). The simulation protocol follows Ref.~\cite{benvegnen2023meta}, where the initial liquid configuration is constructed with particles of $\sigma=3$ (blue) state and thermalized up to a time $t_{\rm th}=150$ to allow the system to adapt to the changed environment.  At $t=0$, a circular region of radius $r_d$, centered at $(L_x/2, L_y/2)$, is chosen, and an additional $\Delta N = (\rho_0^d - \rho_0) \pi r_d^2$ particles are inserted to form a high-density droplet ($\rho_0^d > \rho_0$). Here, $\rho_0^d$ and $r_d$ denote the droplet’s density and radius, respectively. The internal state of all particles in the droplet is then set to $\sigma = 1$ (red) for a counter-propagating droplet, or to $\sigma = 2$ (green) or $\sigma = 4$ (black) for a transversely propagating droplet. Simulations are carried out across a range of control parameters: the diffusion constant $D$, the inverse temperature $\beta = 1/T$ controlling noise, and the self-propulsion strength $\epsilon$, which sets the effective particle speed. The system is evolved using a random-sequential-update Monte Carlo algorithm until a stationary state is reached at $t_{\rm eq} \sim 10^3$--$10^5$, after which ensemble-averaged measurements are recorded.

Fig.~\ref{fig:fig_drop_fate} illustrates the time evolution and resulting steady states following artificial droplet excitations at different self-propulsion velocities, deep inside the liquid phase ($\beta=1$, $\rho_0=10$). As shown in Fig.~\ref{fig:fig_drop_fate}(a--f), a small counter-propagating droplet does not destabilize the polar ordered phase at a low bias of $\epsilon=0.3$ [Fig.~\ref{fig:fig_drop_fate}(a--b)], but it disrupts the liquid phase at intermediate and high biases, $\epsilon=0.9$ and $\epsilon=2.7$ [Fig.~\ref{fig:fig_drop_fate}(c--f)], thereby confirming the metastable character of the polar liquid phase in the 4-state APM. For $\epsilon=0.3$ [Fig.~\ref{fig:fig_drop_fate}(a--b)], the directional hopping rate is low, while the non-directional hopping rates, although smaller, remain appreciable [see Eq.~\eqref{whop}]. As a consequence, the droplet dissolves gradually without perturbing the original ordered phase composed of $\sigma=3$ particles. At $\epsilon=0.9$ [Fig.~\ref{fig:fig_drop_fate}(c--d)], we observe that the droplet affects the ordered background similarly to the behavior seen in the AIM~\cite{benvegnen2023meta}, expanding at the cost of the ordered region and leaving behind a comet-shaped trail as it advances along its preferred direction (see Appendix~\ref{drop_pos} for the time evolution of such a counter-propagating droplet at low and high self-propulsion velocities). This growth results from fluctuations in orientation at the droplet boundary, where blue $\sigma=3$ particles convert into red $\sigma=1$ particles, fueling further expansion. As the transformation proceeds, the original polar phase (blue, $\sigma=3$) is entirely replaced by a new ordered phase associated with the droplet (red, $\sigma=1$), as seen in Fig.~\ref{fig:fig_drop_fate}(d). In contrast, the limited transverse hopping along the $\pm y$-direction at higher propulsion $\epsilon=2.7$ prevents significant transverse expansion of the droplet [Fig.~\ref{fig:fig_drop_fate}(e--f)]. As discussed in the context of Fig.~\ref{fig:apm_ftcs_snaps}, this constraint leads to the development of a partially reversed mixed sandwich configuration [Fig.~\ref{fig:fig_drop_fate}(f)], in which regions of the original ordered phase (blue) coexist with a red lane formed by the longitudinal motion of the droplet. This sandwich structure constitutes a stable steady-state that was not reported in previous studies involving droplet excitations~\cite{codina2022small,benvegnen2023meta}, although such a configuration can emerge for reduced transverse diffusion (see Appendix~\ref{sandwich_aim} for more detail). Importantly, the final position of the lane in Fig.~\ref{fig:fig_drop_fate}(f) depends on the initial placement of the droplet, indicating that multiple stable sandwich steady-states may arise. This structure can also emerge spontaneously in the APM for large bias $\epsilon$, even in the absence of any initial droplet perturbation, when starting from random configurations.

The effect of transversely propagating droplets at low and high self-propulsion velocities is illustrated in Fig.~\ref{fig:fig_drop_fate}(g--j). At low self-propulsion (small $\epsilon$), the comparatively high rate of unbiased hopping results in significant transverse diffusion along the droplet’s boundary. This promotes mixing with the surrounding medium, preventing the droplet from maintaining its internal order. Consequently, the droplet disperses into the stable transversely moving liquid phase, leaving the initial ordered state intact [Fig.~\ref{fig:fig_drop_fate}(g--h)]. Therefore, at low $\epsilon$, irrespective of the droplet’s direction of propagation, strong transverse diffusion inhibits its growth, causing it to dissolve without disrupting the surrounding liquid phase. In contrast, at high self-propulsion (large $\epsilon$), the biased hopping becomes dominant while the unbiased hopping is greatly reduced. As a result, the droplet experiences limited diffusion at its interface while simultaneously advancing more rapidly along its preferred direction, reducing the time it interacts with the ordered background. This directional persistence enables the droplet to avoid disintegration from the background flow and instead expand, ultimately destroying the existing ordered phase [Fig.~\ref{fig:fig_drop_fate}(i--j)].

Although Fig.~\ref{fig:fig_drop_fate}(d) illustrates a full reversal of the initial liquid phase triggered by a counter-propagating droplet, this apparent reversal is in fact a finite-size effect. Let $w_{\rm sw}$ be the sandwich width for the infinite system, then one observes a complete reversal in a finite system when $L_y < w_{\rm sw}$ (keeping $L_x$ fixed). We, therefore, conclude that deep inside the liquid phase (low noise) and in the thermodynamic limit, the sandwich state becomes the unique steady-state outcome following a counter-propagating droplet perturbation. See Appendix~\ref{sandwich} for a detailed discussion of how spatial anisotropy governs the stability of the APM sandwich state. In what follows, we provide a quantitative analysis of how $L_x$, $L_y$, and $\epsilon$ influence the emergence of this configuration.

\begin{figure}[!t]
    \centering
    \includegraphics[width=\columnwidth]{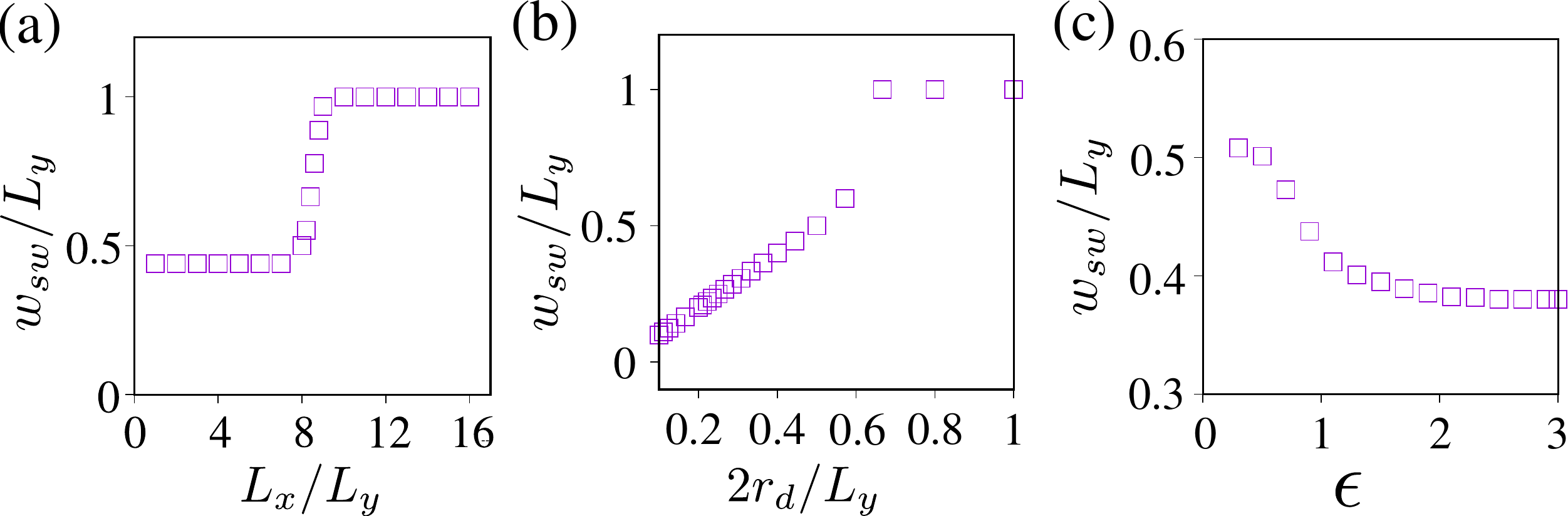}
    \caption{(color online) {\it Phase reversal due to counter-propagating droplet}. (a) Average relative width ($w_{sw}/L_y$) of the sandwich state versus aspect ratio of the simulation box $(L_x/L_y)$ for fixed $r_d = 10$ and $\epsilon=0.9$. $w_{sw}/L_y=1$ corresponds to a full reversal of the initial ordered state. (b) $w_{sw}/L_y$  as a function of relative droplet size $(2r_d/L_y)$ obtained for fixed $r_d = 6$ and $\epsilon=0.9$. (c) $w_{sw}/L_y$ as a function of $\epsilon$ for counter-propagating droplet with fixed $L_y=50$ and $r_d = 10$. A movie (\texttt{movie03}) showing the sandwich formation for various $\epsilon$ can be found at Ref.~\cite{zenodo}. Parameters: $D=1$, $\beta=1$, $\rho_0=10$, $\rho_0^d = 5\rho_0$ and (a) $L_y=50$, (b--c) $L_x = 200$.}
    \label{fig:fig_Lx_Pr_vs_Ly_sw}
\end{figure}

\begin{figure*}[!t]
    \centering
    \includegraphics[width=0.85\textwidth]{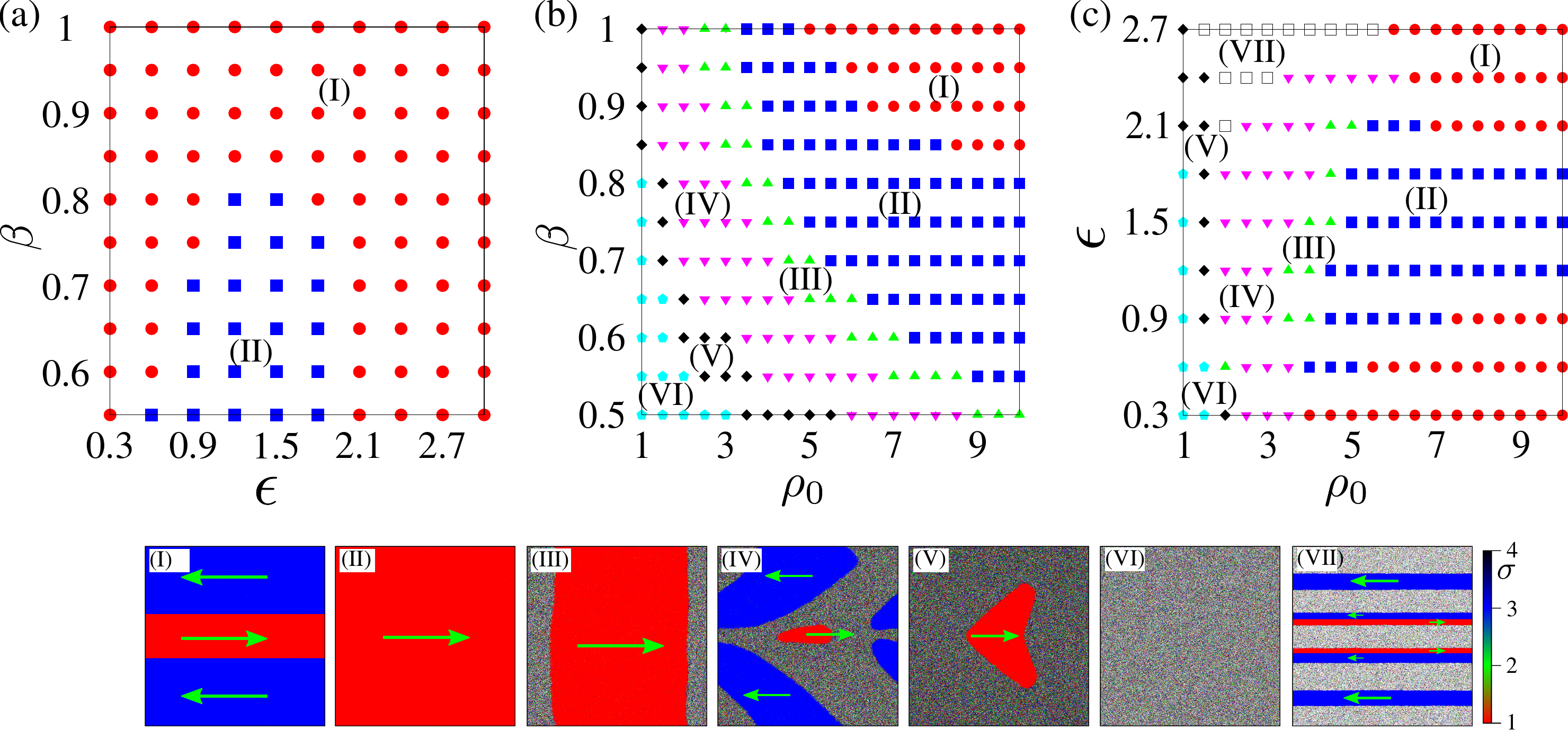}
    \caption{(color online) {\it State diagrams for counter-propagating droplet excitation}. [above] (a) $\beta-\epsilon$ diagram for $\rho_0=10$. (b) $\beta-\rho_0$ diagram for $\epsilon=1.5$. and (c) $\epsilon-\rho_0$ diagram for $\beta=0.75$. Each diagram is partitioned into regions labeled (I)--(VII), corresponding to different steady-states: (I) sandwich state (circle), (II) full reversal of the initial liquid phase by the droplet (square), (III) coexistence band (up triangle), (IV) coexistence sandwich (down triangle), (V) liquid blob of the droplet on a gaseous background (diamond), (VI) gas phase (pentagon), and (VII) longitudinal lane (open square), primarily composed of the initial liquid phase, though it can also form from other states.  Parameters: $D=1$, $r_d = 10$, $\rho_0^d = 5\rho_0$, and $L_x = L_y = 200$. [below] Representative snapshots of the seven steady states in the phase diagram for $L_x = L_y = 800$.}
    \label{fig:fig_PD_counter}
\end{figure*}

Fig.~\ref{fig:fig_Lx_Pr_vs_Ly_sw} illustrates how spatial anisotropy and self-propulsion strength influence the reversal dynamics of the flocking phase under the action of counter-propagating droplets. In Fig.~\ref{fig:fig_Lx_Pr_vs_Ly_sw}(a), the relative width of the sandwich region, $w_{\rm sw}/L_y$, is plotted against the aspect ratio of the simulation domain at a fixed transverse size $L_y=50$. As $L_x$ increases, the system undergoes a transition from a sandwich configuration $(w_{\rm sw}/L_y<1)$ to a full reversal $(w_{\rm sw}/L_y=1)$ of the original ordered phase. This behavior arises because a larger longitudinal size allows the droplet more time to grow and move before its front encounters its trailing end, enabling it to span the entire system vertically. In contrast, for smaller $L_x$, the growth of the droplet front is interrupted by its own trajectory, leading to a persistent sandwich state. Fig.~\ref{fig:fig_Lx_Pr_vs_Ly_sw}(b) shows that, for fixed droplet radius $r_d=6$, the sandwich width $w_{\rm sw}/L_y$ exhibits a discontinuous change as a function of $L_y$. Complete reversal is favored when $L_y$ is close to the droplet diameter, but becomes impossible as $L_y$ grows significantly beyond it. In this regime, the width of the sandwich region $w_{\rm sw}$ becomes independent of $L_y$, leading to a linear increase of $w_{\rm sw}/L_y$ with $1/L_y$. Thus, full reversal is favored only at small transverse sizes. For large self-propulsion $\epsilon$, this linear trend persists even for smaller $L_y$, and full reversal occurs only when $L_y \approx 2r_d$. In Fig.~\ref{fig:fig_Lx_Pr_vs_Ly_sw}(c), $w_{\rm sw}/L_y$ is plotted against $\epsilon$ for $L_y=50$. At low $\epsilon$, transverse diffusion is significant, resulting in $w_{\rm sw} > 2r_d$. As $\epsilon$ increases, transverse hopping is suppressed, and $w_{\rm sw}/L_y$ decreases gradually until around $\epsilon \sim 1.4$, the point at which the system transitions from transverse to longitudinal motion~\cite{chatterjee2020flocking,mangeat2020flocking}. Beyond $\epsilon \gtrsim 2$, transverse motion becomes negligible, and the sandwich width saturates at approximately $w_{\rm sw} \simeq 2r_d - 1$. Across this range, the sandwich state emerges as the generic steady state for counter-propagating droplet excitations. This trend is similar to the one obtained from the hydrodynamic theory shown in Fig.~\ref{fig:apm_ftcs_snaps}(f).

Additionally, we report in Appendix~\ref{sw_stability} that the sandwich state is not density-segregated. The density profile, integrated over the x-coordinate, remains uniform; only the internal polarization changes along the transverse direction. Appendix~\ref{sw_stability} also presents a stability diagram of the sandwich state, identifying that this state remains stable only if the initial width is between two thresholds: $w_{\min} < w_{\rm sw}< w_{\max}$, otherwise the system relaxes to a liquid phase of a single state. Therefore, for a given set of control parameters $(D, \beta, \rho_0)$, one can prepare the system either in the uniform ordered flocking state or in a sandwich state with any width in the interval $\in [w_{\min}(\epsilon),w_{\max}(\epsilon)]$, and all such states are stable. In this sense, the model admits a continuous family of stable sandwich states with distinct widths. A localized droplet perturbation applied to the ordered flocking state just drives the system into one member of this family, selecting a final width within the stability window.

Results from the APM hydrodynamic theory [Sec.~\ref{sec:hydro}] indicate that the system exhibits qualitatively different responses to counter-propagating droplet excitations depending on the noise strength [Fig.~\ref{fig:apm_ftcs_snaps}(a--b)]. At low noise, the droplet stabilizes into a sandwich configuration, whereas at high noise, it can fully destabilize the background liquid phase. We have already discussed in length the low-noise regime deep inside the liquid phase, where the sandwich state is the dominant and stable outcome. Now, we return to the high-noise regime to demonstrate that for stronger noise strength $(\beta \leq 0.82)$, the stability of the sandwich state weakens, and a counter-propagating droplet can completely reverse the initial liquid phase, in line with the hydrodynamic prediction (see Fig.~\ref{fig:apm_ftcs_snaps}(b), Appendix~\ref{CD_beta} and \texttt{movie04} in Ref.~\cite{zenodo} for further details). In Fig.~\ref{fig:fig_PD_counter}, we systematically explore the response of the system with counter-propagating droplet excitation by constructing comprehensive phase diagrams across the full range of control parameters ($\beta$, $\epsilon$, and $\rho_0$). The $\beta-\epsilon$ diagram [Fig.~\ref{fig:fig_PD_counter}(a)] is shown for a very high density $(\rho_0=10)$ and dominated by the sandwich configuration (I) when noise is small and the system is deep inside the liquid phase. However, as noise increases, the system undergoes complete reversal (II) for intermediate values of self-propulsion, where the counter-propagating droplet progressively displaces and replaces the initial liquid phase. Both the $\beta-\rho_0$ and $\epsilon-\rho_0$ diagrams [Figs.~\ref{fig:fig_PD_counter}(b--c)] reveal a rich variety of steady states as the system moves away from the liquid phase, either by increasing noise or decreasing density. These include a high-density band of the droplet state forming a liquid-gas coexistence regime after destroying the initial liquid (III), a sandwich configuration in the coexistence regime (IV), a large liquid blob of the droplet state (V), pure gas phase (VI), and longitudinal lane formation at higher $\epsilon$ (VII). While the phase boundaries in the diagrams are obtained for a finite system size ($L = 200$) and may shift slightly with increasing system size, the overall structure and classification of steady states remain robust. To demonstrate that these phases are not finite-size artifacts, we provide representative snapshots of the steady-states for a larger system size ($L = 800$), confirming the persistence and stability of the identified states in the thermodynamic limit.

In the case of a droplet excitation moving transversely, the resulting steady-state configuration also depends sensitively on the particle self-propulsion velocity. At intermediate values of $\epsilon$, the system predominantly exhibits a morphology where two distinct high-density clusters coexist, each maintaining persistent motion in perpendicular directions—one representing the inserted droplet and the other corresponding to the preexisting liquid phase (see Fig.~\ref{fig:apm_ftcs_snaps}(d), Appendix~\ref{CD_beta} and \texttt{movie05} at Ref.~\cite{zenodo} for $0.9 \leq \epsilon \leq 1.5$). These clusters, though they frequently collide, do not merge into a single homogeneous phase, highlighting a dynamical mismatch in their collective movement. At higher self-propulsion strengths, however, the transverse droplet fully destabilizes the initial liquid, and the system transitions into a final state comprised entirely of particles in the droplet's spin state, or into a mixed phase containing both spin types that support transverse motion with respect to the direction of the original liquid flow (see \texttt{movie05} at Ref.\cite{zenodo} for $\epsilon \geq 2.4$).

\begin{figure}[!t]
    \centering
    \includegraphics[width=\columnwidth]{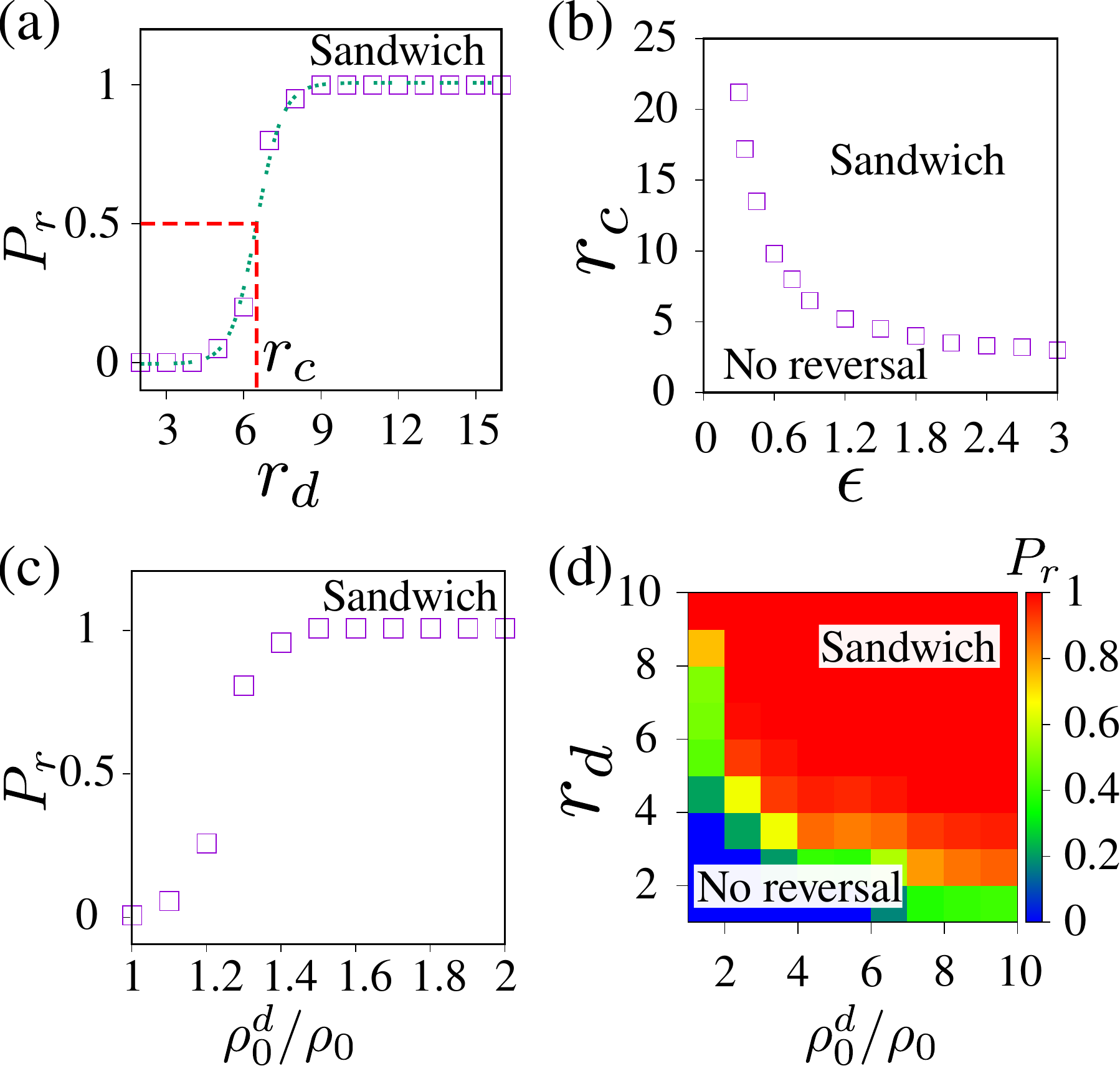}
    \caption{(color online) {\it Liquid stability with a counter-propagating droplet}. (a) Sandwich state probability $P_r$ as a function of droplet radius $r_d$ for fixed $\epsilon=0.9$ and $\rho_0^d = 1.2\rho_0$. The dotted line is a hyperbolic tangent fit for extracting the critical radius $r_c$ when $P_r=0.5$. (b) $r_c$ versus $\epsilon$ at a fixed initial droplet density $\rho_0^d = 1.2\rho_0$. (c) The probability of sandwich state, $P_r$, as a function of $\rho_0^d$ for fixed $r_d=5$ and $\epsilon=1.2$. (d) $(\rho_0^d,r_d)$ stability diagram for $\epsilon=0.9$. The color bar denotes the sandwich state probability, $P_r$. Parameters: $D=1$, $\beta=1$, $\rho_0=10$, and $L_x = L_y = 200$.}
    \label{fig:fig_Pr_r_rc_epsilon}
\end{figure}

Thus far, we have shown that droplet insertion into the liquid phase of the APM can destabilize it, resulting either in a sandwich state or a complete reversal of the ordered phase. We now provide a quantitative analysis of how the size and density of the droplet influence this reversal. Figs.~\ref{fig:fig_Pr_r_rc_epsilon}(a,c) display the probability of sandwich formation, $P_r$, for counter-propagating droplets as functions of droplet radius $(r_d)$ and droplet density $(\rho_0^d)$, respectively. In Fig.~\ref{fig:fig_Pr_r_rc_epsilon}(a), the dashed line shows a fit using a hyperbolic tangent function~\cite{benvegnen2023meta}, which is used to identify the critical radius, $r_c$, marking the transition from the initial ordered phase to the sandwich state triggered by a counter-propagating droplet. The self-propulsion velocity, $\epsilon$, also significantly affects the phase stability. Fig.~\ref{fig:fig_Pr_r_rc_epsilon}(b) shows how the critical radius $r_c$ varies with $\epsilon$. At higher bias, smaller droplets are sufficient to generate a sandwich state, due to enhanced hopping in the preferred direction and suppressed transverse movement. As $\epsilon$ decreases and transverse diffusion becomes more prominent, a larger $r_c$ is needed to achieve the transition. The phase diagram in Fig.~\ref{fig:fig_Pr_r_rc_epsilon}(d), plotted in the $(r_d, \rho_0^d)$ plane, indicates that the sandwich state remains the dominant steady-state configuration with counter-propagating droplet excitation. This phase diagram resembles the one obtained from the hydrodynamic theory shown in Fig.~\ref{fig:apm_ftcs_snaps}(g).

\begin{figure}[!t]
    \centering
    \includegraphics[width=\columnwidth]{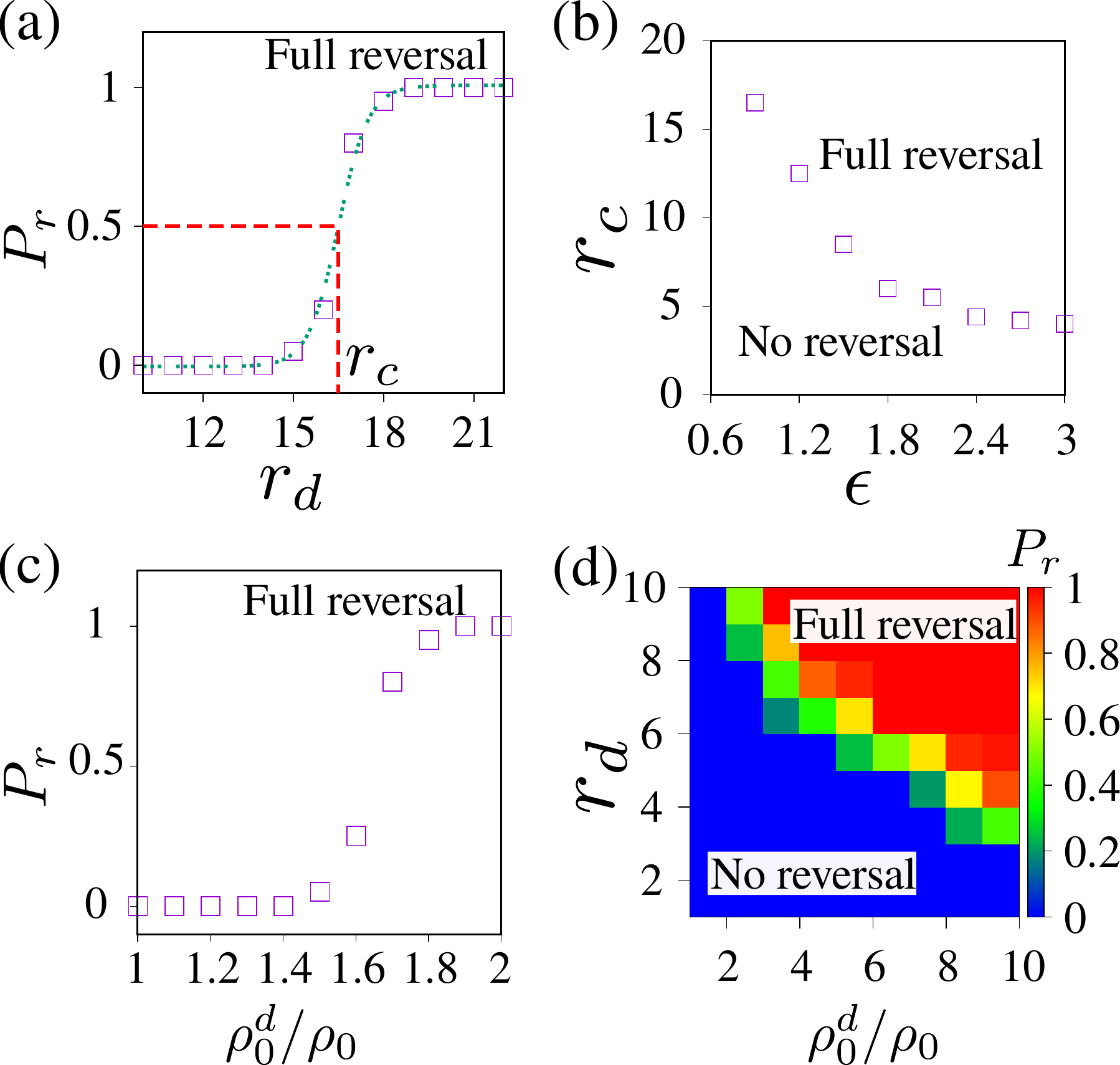}
    \caption{(color online) {\it Liquid stability with a transversely-propagating droplet}. (a) Probability of full reversal of the initial liquid phase, $P_r$, for fixed $\epsilon=0.9$ and $\rho_0^d = 1.2\rho_0$. The dotted line is fit to a hyperbolic tangent used to extract $r_c$. (b) The critical radius, $r_c$, as a function of $\epsilon$ for fixed $\rho_0^d = 1.2\rho_0$. (c) The probability of a full reversal of the ordered phase, $P_r$, as a function of the initial droplet density $\rho_0^d$ for fixed $r_d=5$ and $\epsilon=1.2$. (d) $(\rho_0^d,r_d)$ stability diagram for $\epsilon=0.9$. The color bar denotes the probability of a full reversal, $P_r$. Parameters: $D=1$, $\beta=1$, $\rho_0=10$, and $L_x = L_y = 200$.}
    \label{fig:fig_Rad_vs_Frac}
\end{figure}

Fig.~\ref{fig:fig_Rad_vs_Frac}, which corresponds to the case of transverse droplet propagation, is analogous to Fig.~\ref{fig:fig_Pr_r_rc_epsilon} for counter-propagating droplets. It presents how transverse motion relative to the homogeneous liquid phase affects phase reversal. An increase in the initial droplet density $\rho_0^d$ tends to stabilize the droplet, increasing the likelihood of a complete reversal of the ordered phase. Similarly, a larger droplet radius $r_d$ enhances the probability of reversal, indicating that both parameters play key roles in determining the stability of the phase. Nonetheless, compared to counter-propagating droplets, a stronger perturbation is needed to achieve complete reversal in the transverse case [Figs.~\ref{fig:fig_Rad_vs_Frac}(a,c)]. A comparison between Figs.~\ref{fig:fig_Rad_vs_Frac}(a,c) and Figs.~\ref{fig:fig_Pr_r_rc_epsilon}(a,c) shows that counter-propagating droplets can destabilize the ordered phase more efficiently, requiring smaller $r_d$ and lower $\rho_0^d$ than their transversely propagating counterparts. In the case of transverse propagation, the critical radius $r_c$ necessary for full reversal increases sharply as $\epsilon$ decreases, unlike in the counter-propagating scenario [Fig.~\ref{fig:fig_Rad_vs_Frac}(b)]. At small $\epsilon$, enhanced non-biased hopping leads to greater diffusion at the droplet boundary, causing the transversely moving droplet to dissipate. As a result, a significantly larger droplet is needed to induce full reversal when $\epsilon$ is low. The stability diagram in Fig.~\ref{fig:fig_Rad_vs_Frac}(d), plotted in the $(\rho_0^d, r_d)$ plane, further characterizes the behavior for transverse droplet propagation. The fully reversed state, when it occurs, features two high-density clusters moving in perpendicular directions, corresponding to the droplet and the initial liquid, as described in Appendix~\ref{CD_beta}. This phase diagram is similar to the one obtained from the hydrodynamic theory shown in Fig.~\ref{fig:apm_ftcs_snaps}(h). All these findings with a droplet excitation are then consistent with the results obtained from the hydrodynamic analysis presented in Sec.~\ref{sec:hydro}.

\begin{figure}[!t]
    \centering
    \includegraphics[width=\columnwidth]{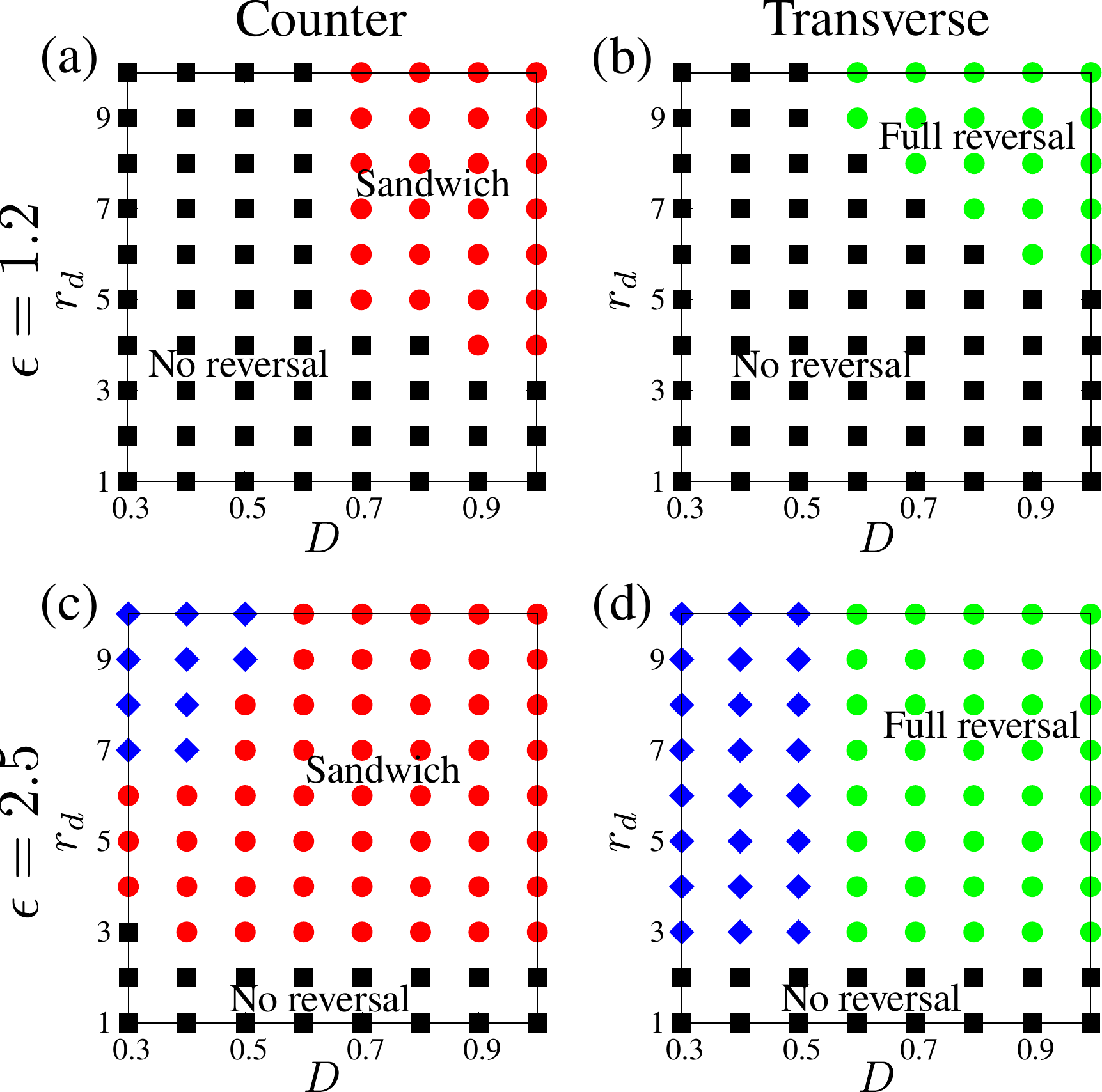}
    \caption{(color online) {\it Liquid stability with droplet excitations for varying diffusion}. $D-r_d$ stability diagrams for (a--b) $\epsilon=1.2$ and (c--d) $\epsilon=2.5$ for (a, c) counter and (b, d) transversely-propagating droplets. Parameters: $\beta=1$, $\rho_0=10$, $\rho_0^d=1.5\rho_0$, and $L_x = L_y = 200$.}
    \label{fig:fig_rd_D_phasediagram}
\end{figure}

So far, we have reported results for artificial droplet excitation in the APM at a diffusion constant $D=1$, consistent with earlier studies of the model~\cite{chatterjee2020flocking, mangeat2020flocking}, which primarily focused on steady states at this representative value. However, recent investigations~\cite{benvegnen2023meta,jd2024MIP,TSAIM} have demonstrated that lower diffusion can substantially influence the emergent collective dynamics in discrete flocking systems. This motivates a more systematic analysis of diffusion’s role in our model, as presented in Fig.~\ref{fig:fig_rd_D_phasediagram}, which shows four stability diagrams characterizing the droplet-induced behavior of the APM for reduced diffusion. For $\epsilon=1.2$ [Fig.~\ref{fig:fig_rd_D_phasediagram}(a--b)], the diagrams indicate that achieving either a sandwich state [Fig.~\ref{fig:fig_rd_D_phasediagram}(a)] or a full reversal [Fig.\ref{fig:fig_rd_D_phasediagram}(b)] of the initial liquid phase requires higher diffusion as the droplet radius decreases. At this value of $\epsilon$, hopping in non-preferred directions remains appreciable, meaning that only larger droplets can persist when $D$ is small; smaller droplets dissipate over time. As a result, for $D \leq 0.5$, no droplets survive, regardless of their size (up to the maximum tested $r_d=10$). Notably, in Fig.\ref{fig:fig_rd_D_phasediagram}(b), the full reversal state observed at $D \geq 0.7$ and larger $r_d$ $(r_d \gtrsim 7)$ is marked by orthogonal motion of clusters associated with the droplet and the initial liquid, as detailed in Appendix\ref{CD_beta}.

For larger $\epsilon$ [Fig.~\ref{fig:fig_rd_D_phasediagram}(c--d)], the suppressed hopping in non-preferred directions allows droplets with smaller radii to persist, making both the sandwich state and complete reversal achievable for higher diffusion values $(D \geq 0.6)$. However, at lower diffusion levels $(0.3 \leq D \leq 0.5)$—indicated by blue diamonds—a droplet (in state $\sigma_d=1$ or $\sigma_d=2$) fully disrupts the original liquid phase (state $\sigma = 3$), giving rise to a dense lane consisting of particles in its own spin state. Over time, however, this lane becomes unstable due to the spontaneous emergence of another transversely moving spin state, $\sigma_d \pm 1[\mathrm{mod}4]$, within it. This newly nucleated spin state then displaces the existing lane and forms a new one. The process iterates, with each newly established lane being replaced by another, driven by spontaneous nucleation of transversely polarized droplets. This dynamic cycling of lane creation and annihilation, governed by competing transverse spin states, has also been observed in Refs.~\cite{jd2024MIP,TSAIM}. An illustration of this behavior for $D=0.3$ is provided in \texttt{movie06} at Ref.~\cite{zenodo}.

For $D = 0.1$ and $D = 0.2$, the system exhibits qualitatively different behavior compared to earlier cases: the inserted droplet forms a jammed cluster, even in the absence of external constraints on particle motion, distinct from the jamming observed in Ref.~\cite{karmakar2023jamming}. At $D = 0.2$, this jammed cluster gradually dissolves as its constituent particles flip to match the spin of the background liquid, ultimately restoring the initial ordered phase (see \texttt{movie07} at Ref.~\cite{zenodo}). In contrast, at $D = 0.1$, low motility combined with reduced noise ($\beta = 1$) gives rise to the spontaneous nucleation of oppositely polarized droplets outside the original droplet, leading to the formation of multiple jammed domains. These domains appear either as horizontal clusters of $\sigma = 2$ (lower side) and $\sigma = 4$ (upper side), or as vertical clusters of $\sigma = 1$ (left) and $\sigma = 3$ (right). The extent and configuration of these jammed regions, with sharply polarized domains, are governed by the degree of self-propulsion and act as effective barriers that hinder subsequent particle motion and rearrangement (see \texttt{movie08} at Ref.~\cite{zenodo}).

Fig.~\ref{fig:fig_rd_D_phasediagram}(c--d) highlights how low diffusion and strong self-propulsion facilitate the spontaneous emergence of transversely polarized droplets within the high-density phase of the APM, which ultimately destabilize and fragment the original liquid structure. Motivated by this, we now turn to a more detailed exploration of the APM under conditions of low diffusion and high self-propulsion.


\subsection{Numerical simulations in the regime of small diffusion}
\label{sec:sponnucl}

\begin{figure*}
    \centering
    \includegraphics[width=1.5\columnwidth]{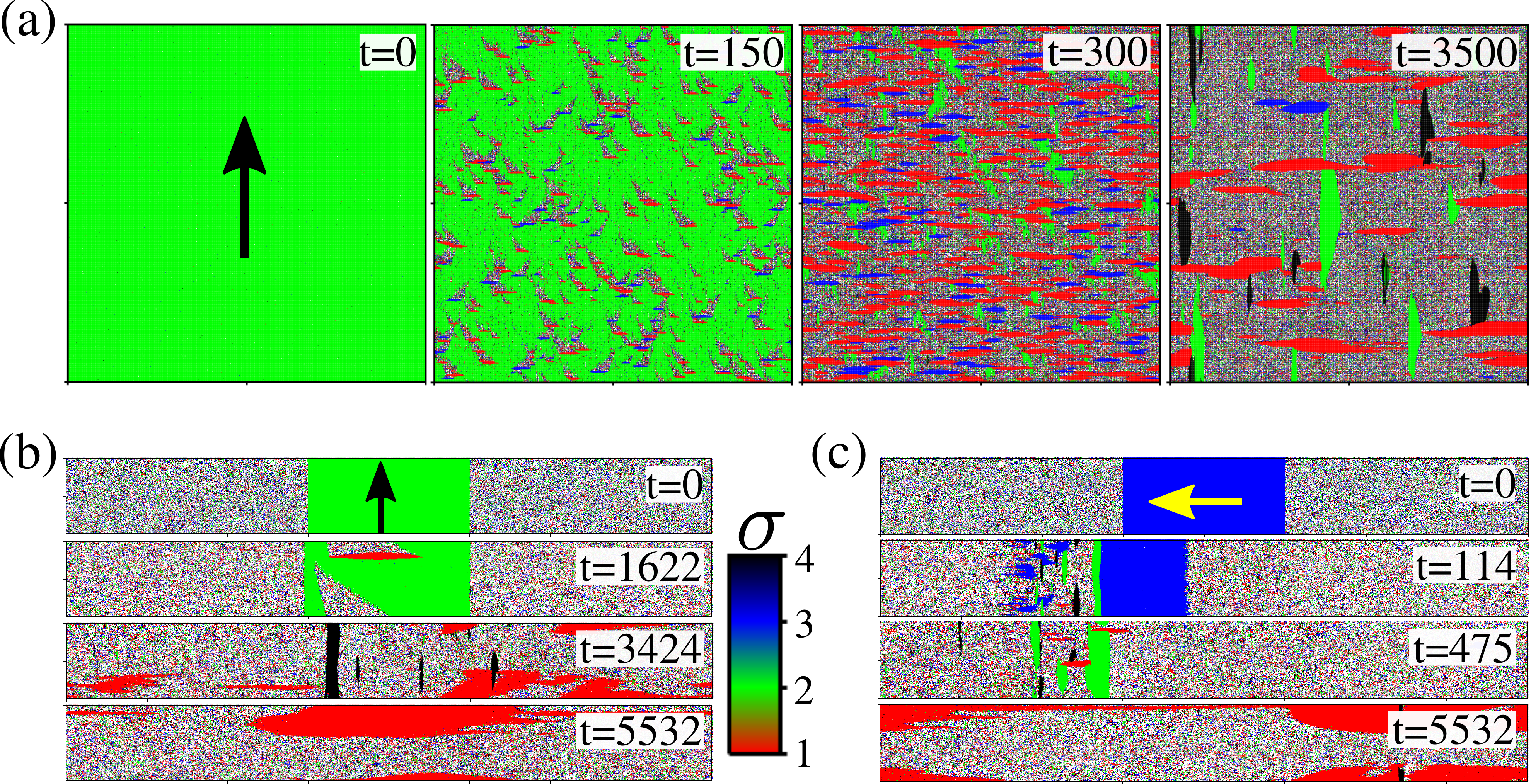}
    \caption{(color online) {\it Spontaneous nucleation of polar droplets in the liquid phase}. Time evolution snapshots after starting from (a) ordered ($\sigma=2$ particles, green) and (b--c) semi-ordered configurations [(b) longitudinal band of $\sigma=2$ particles, (c) transverse band of $\sigma=3$ particles] exhibit spontaneous nucleation of droplets of other spin states and subsequent destruction of the long-range ordered (LRO) phase. Parameters: $D=0.3$, $\beta=1$, $\epsilon=2.5$, (a) $\rho_0=5$ and (b--c) $\rho_0=3$. System size: (a) square domain of dimension $1024 \times 1024$ and (b--c) rectangular domain of dimension $1024 \times 128$. A movie (upper panel of \texttt{movie09}) of the same can be found at Ref.~\cite{zenodo}.}
    \label{fig:fig_spontaneous_nucleation}
\end{figure*}

Recent investigations of the one-species AIM~\cite{benvegnen2023meta,jd2024MIP} and the two-species AIM with both reciprocal and non-reciprocal inter-species interactions~\cite{TSAIM} have shown that, at low diffusion, spontaneous nucleation of droplets with opposite polarity can destabilize the flocking phase. This leads to a breakdown of global order, resulting in multiple clusters with only short-range correlations. Furthermore, it has been reported~\cite{benvegnen2023meta} that the nucleation time for such droplets scales as $\sim e^D / L^2$, where $L$ denotes the linear system size. In Sec.~\ref{sec:droplet}, we similarly observed the spontaneous emergence of droplets within the dense liquid phase of the APM, propagating transversely to the main flow, for low diffusion and strong bias. Motivated by these findings, we now analyze the 4-state APM in the absence of any externally inserted droplet, focusing on regimes of small diffusion and high self-propulsion.

Fig.~\ref{fig:fig_spontaneous_nucleation} illustrates the spontaneous nucleation of droplets composed of particles biased to move transversely to the liquid flow, which ultimately destabilize both the polar liquid and the liquid-gas coexistence band, consistent with earlier reports for the AIM~\cite{jd2024MIP}. Starting from a homogeneous liquid of $\sigma=2$ particles [Fig.~\ref{fig:fig_spontaneous_nucleation}(a), $t=0$], left- $(\sigma=3)$ and right-moving $(\sigma=1)$ clusters emerge spontaneously by $t=150$. These clusters grow over time $(t=300)$, alongside new spin states, and progressively destroy the long-range order (LRO) of the initial phase. At $t=3500$, the system settles into a steady state with short-range order (SRO), composed of multiple mobile domains of all four spin states. Due to the large self-propulsion bias $(\epsilon = 2.5)$, these clusters align longitudinally with their direction of motion.

This instability also affects the coexistence phase. Figs.~\ref{fig:fig_spontaneous_nucleation}(b--c) show that both transversely and longitudinally moving bands eventually destabilize into SRO phases, though the nucleation time differs: transverse bands break down more rapidly ($t = 114$) than longitudinal ones ($t = 1622$). This is expected, as the APM favors longitudinal band motion at high $\epsilon$ due to the reorientation transition~\cite{chatterjee2020flocking, mangeat2020flocking}. In both cases, the final SRO state is qualitatively similar at long times ($t = 5532$).

Unlike the AIM~\cite{jd2024MIP}, where nucleation predominantly proceeds via counter-propagating droplets, the APM exhibits spontaneous nucleation exclusively via transversely propagating droplets, which then grow and disrupt the ordered phase. This distinction arises from the orthogonal alignment of such droplets relative to the background flow, reducing head-on collisions and enhancing persistence. At low diffusion, stochastic spin fluctuations are suppressed, allowing local spin-aligned regions to survive and coarsen through ferromagnetic alignment. The large bias $\epsilon$ further amplifies the directed motion of these droplets, enabling them to evade dominant background flow and resist dissolution. In contrast, counter-propagating droplets face stronger mechanical resistance and are more prone to decay. As discussed in Sec.~\ref{sec:droplet}, even at large $\epsilon$, counter-propagating droplets tend to produce sandwich states rather than induce full phase reversal, whereas transverse droplets are more effective in destabilizing the ordered phase.

We have also studied the spontaneous nucleation and breakdown of the liquid phase in a modified version of the APM, which includes a small change to the hopping mechanism inspired by the AIM variant proposed in Ref.~\cite{jd2024MIP}. Details of this extension are provided in Appendix~\ref{newAPM}.

\begin{figure}[!t]
    \centering
    \includegraphics[width=\columnwidth]{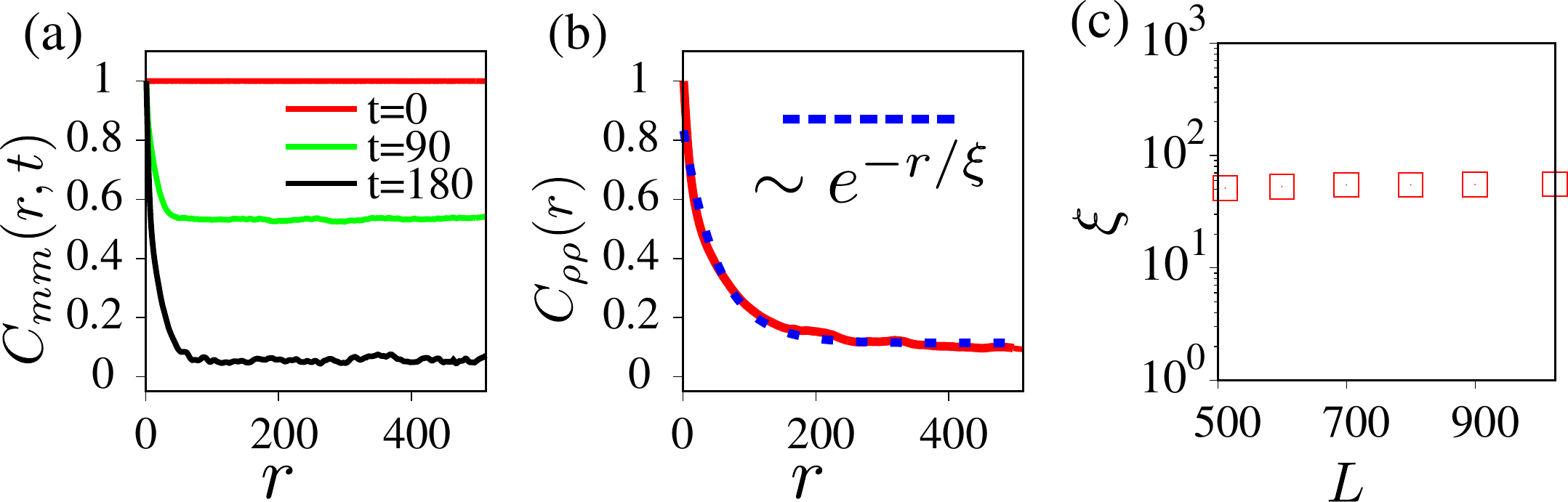}
    \caption{(color online) {\it Correlation length of the SRO state}. (a) Two-point equal time magnetization correlation function $C_{mm}(r,t)$ exhibiting an LRO to SRO transition with time. (b) Exponential decay of the two-point equal time density correlation function $C_{\rho\rho}(r)$ with distance $r$. The blue dotted line represents an exponential fit, estimating a correlation length of $\xi = 51$. System size: $L = 1024$. (c) The correlation length $\xi$ (red square) is independent of system size $L$. Parameters: $D=0.3$, $\beta=1$, $\epsilon=2.5$, and $\rho_0=5$.}
    \label{fig:fig_corr_tnucl}
\end{figure}

We now examine the SRO steady state that emerges due to spontaneous droplet nucleation in the APM liquid phase and estimate the typical cluster sizes as shown in Fig.~\ref{fig:fig_corr_tnucl}. Figs.~\ref{fig:fig_corr_tnucl}(a--b) respectively display an exponential decay in the magnetization and density correlation functions over time, suggesting that the polar flock transitions into a new steady state with spatially disordered local domains, indicating short-range order resulting from droplet nucleation. Fig.~\ref{fig:fig_corr_tnucl}(a) illustrates the time evolution of the two-point equal-time magnetization correlation function, $C_{mm}(r,t)=\sum_i m_i^{\sigma=2}(t)m_{i+r}^{\sigma=2}(t)/\rho_0^2L_xL_y$, where initially $C_{mm}(r) \sim \text{const.}$ at $t=0$, and gradually transitions to $C_{mm}(r) \sim \exp(-r/\xi)$ as time increases, with $\xi$ representing the correlation length. The typical size of the short-range clusters, $\xi=51$, is estimated from the exponential decay of the density correlation function, $C_{\rho\rho}(r)=\sum_i \langle \rho_i \rho_{i+r} \rangle/\rho_0^2L_xL_y$, with spatial separation $r$, where $\langle \dots \rangle$ indicates a steady-state temporal average [Fig.~\ref{fig:fig_corr_tnucl}(b)]. This value of $\xi$ is found to be independent of the system size [Fig.~\ref{fig:fig_corr_tnucl}(c)], consistent with observations in the AIM model~\cite{jd2024MIP}.

\begin{figure}[!t]
    \centering
    \includegraphics[width=\columnwidth]{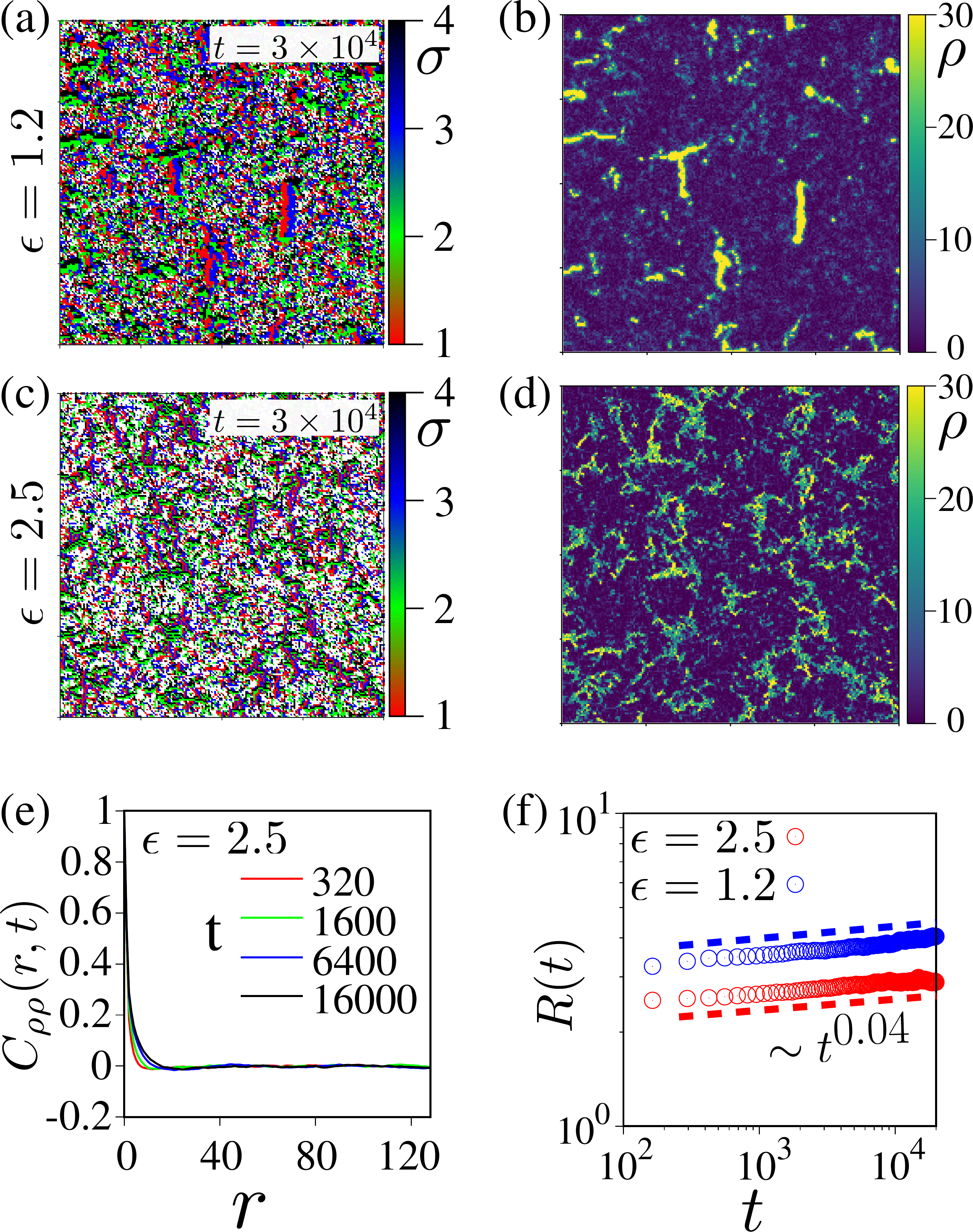}
    \caption{(color online) {\it Motility-induced (interface) pinning in the APM}. (a, c) Starting from an ordered configuration of $\sigma=2$ particles, the system forms jammed clusters in the steady state. The corresponding density plots are shown in (b) and (d). (e) Temporal evolution of the density correlation function $C_{\rho\rho}(r,t)$ (averaged over five ensembles) during the system's transition to MIP. (f) $R(t)$ versus $t$ (on a log-log scale) showing the slow coarsening of the MIP clusters. Parameters: $D=0.3$, $\beta=2$, $\rho_0=5$, and $L_x = L_y = 256$. A movie (\texttt{movie10}) of the same can be found at Ref.~\cite{zenodo}.}
    \label{fig:fig_mip_evo}
\end{figure}

Given that one- and two-species AIM~\cite{jd2024MIP,TSAIM} display motility-induced (interface) pinning (MIP), we investigate the emergence of similar behavior in the APM under conditions of large $\beta$ (low temperature) and reduced diffusion. Initiating from an ordered configuration of $\sigma=2$ particles (green), the system evolves into a steady state featuring jammed clusters [Figs.~\ref{fig:fig_mip_evo}(a--d)], indicating the occurrence of interface pinning at low temperature, consistent with previous findings in the AIM~\cite{jd2024MIP}. At early stages, spontaneously nucleated droplets of opposite spin states (mainly $\sigma=1$ and $\sigma=3$) appear within the liquid region, serving as pinning centers. Consequently, the system rapidly undergoes an MIP transition, where these small droplets progressively aggregate into several narrow, dense pinned clusters over time.

Figs.~\ref{fig:fig_mip_evo}(a--d) depict the late-time MIP morphology in terms of particle state $\sigma$ [Fig.~\ref{fig:fig_mip_evo}(a, c)] and the corresponding density profile [Fig.~\ref{fig:fig_mip_evo}(b, d)] under conditions of low temperature $(\beta=2)$ and small diffusion $(D=0.3)$. A detailed analysis of the prevailing pinned clusters indicates that they represent jammed configurations bordered by sharp interfaces. These jammed regions can orient either vertically or horizontally: in a vertically (horizontally) jammed cluster, the left (top) portion of the interface comprises $\sigma = 1$ $(\sigma=4)$ particles moving to the right (downward), while the right (bottom) side consists of $\sigma = 3$ $(\sigma=2)$ particles moving leftward (upward). Because these spin states move in opposing directions, their interaction causes mutual blockage, giving rise to the jammed structure. Once a particle enters the cluster and crosses the first interface, it travels through the region until reaching the second interface, where it undergoes a spin flip and reverses its direction of motion, effectively mimicking reflection. The particle then moves back toward the first interface, where another spin flip may occur, resulting in continued back-and-forth motion inside the cluster. This mechanism confines particles within the cluster, elevating the local density. Nonetheless, due to the difficulty of escape, the spatial expansion of these jammed regions remains limited. Additionally, the morphology of the jammed clusters depends on the level of self-propulsion: lower self-propulsion yields fewer but broader high-density clusters [Fig.~\ref{fig:fig_mip_evo}(b)], while higher self-propulsion generates many narrower jammed clusters [Fig.~\ref{fig:fig_mip_evo}(d)].

The emergence of the MIP transition at small diffusion can be understood through the development of distinct timescales~\cite{jd2024MIP}. The overall particle hopping rate scales as $4D$, and thus becomes increasingly suppressed as $D$ decreases, whereas spin-flipping processes scale as $\sim \exp(4\beta)$ and become dominant at low temperatures. Additionally, for large $\epsilon$, particle motion becomes persistent rather than diffusive. Consequently, once a jammed cluster forms, marked by the accumulation of oppositely polarized particles on either side of an interface, particles of similar polarity continue to accumulate on their respective sides due to persistent propulsion. These clusters stabilize through pinned interfaces, supported by rapid flipping and suppressed diffusion, which inhibit particle escape and strengthen the phase separation. To clearly illustrate the microscopic mechanism of pinning, we examine the behavior of a counter-propagating droplet. In Appendix~\ref{dropletBETA}, we show how interface pinning of such a droplet can arise under conditions of low temperature and small $D$.

MIP is characterized using the two-point equal-time density correlation function $C_{\rho\rho}(r,t)=\sum_i \delta\rho_i(t) \delta\rho_{i+r}(t) /\rho_0^2 L_x L_y$, where $\delta\rho_i = \rho_i -\rho_0$, and $r$ is the spatial separation at time $t$. As shown in Fig.~\ref{fig:fig_mip_evo}(e), $C_{\rho\rho}(r,t)$ decays rapidly as the initially ordered phase disintegrates and MIP clusters begin to emerge. The time evolution of $C_{\rho\rho}$ reflects a very slow coarsening of MIP domains. To quantify this, we extract a characteristic length scale $R(t)$ representing domain growth. This length scale is defined as the distance at which $C_{\rho\rho}(r,t)$ drops to 30\% of its maximum value. We observe that although $R(t)$ grows over time following a power-law $R(t) \sim t^{1/z}$, the growth is extremely slow and yields an exponent $z \simeq 25$ independent of the strength of self-propulsion. However, at a given time $t$, larger domain size $R(t)$ is observed for smaller $\epsilon$ ($\epsilon = 1.2$), while higher $\epsilon$ ($\epsilon = 2.5$) leads to smaller domain [Fig.~\ref{fig:fig_mip_evo}(f)], consistent with the snapshots in Fig.~\ref{fig:fig_mip_evo}(a--b).

\begin{figure*}[!t]
    \centering
    \includegraphics[width=1.7\columnwidth]{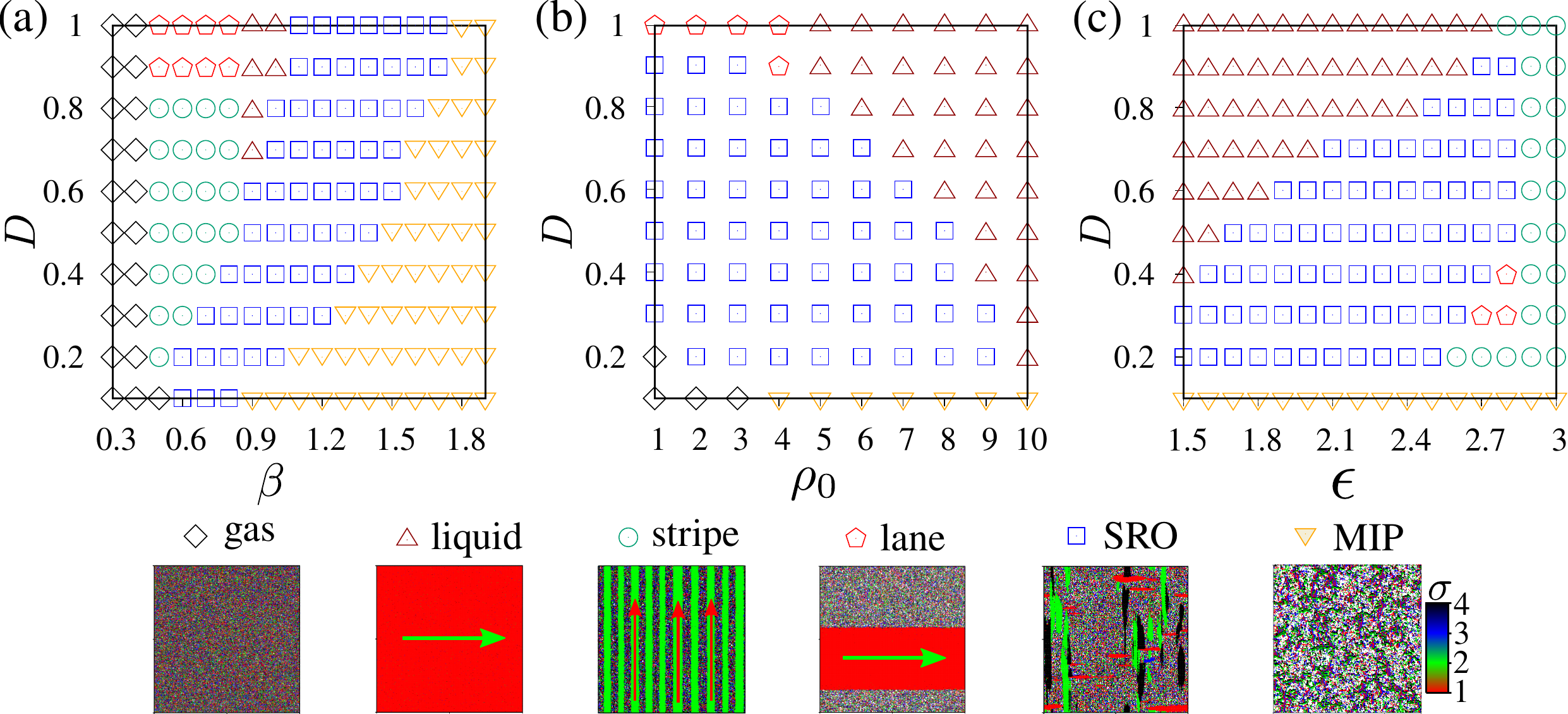}
    \caption{(color online) {\it APM phase diagrams for varying diffusion.} The phase diagrams are shown in: (a) the ($D$--$\beta$) plane for $\rho_0=5$ and $\epsilon=2.5$; (b) the ($D$--$\rho_0$) plane for $\beta=1$ and $\epsilon=2.5$; and (c) the ($D$--$\epsilon$) plane for $\beta=1$ and $\rho_0=5$. Six distinct states are shown in the phase diagrams: gas (black diamond), long-range ordered liquid (dark-red up triangle), stripe (green circle), lane (red pentagon), short-range order (SRO, blue square), and motility-induced pinning (MIP, orange down triangle). System size: $L_x = L_y = 512$.}
    \label{fig:fig_D_vs_Rho_Eps_phase_diagram}
\end{figure*}

To collate the emerging features of the APM as a function of the diffusion constant $D$ and other key parameters (average density $\rho_0$, particle self-propulsion velocity $\epsilon$, and inverse temperature $\beta$), we construct three phase diagrams shown in Fig.~\ref{fig:fig_D_vs_Rho_Eps_phase_diagram}. For $D=1$, we recover the APM behavior reported in Ref.~\cite{chatterjee2020flocking,mangeat2020flocking}, where increasing either the density or $\beta$ drives the system from a coexistence phase (longitudinal lane) to a polar liquid phase exhibiting long-range order (LRO). In Fig.~\ref{fig:fig_D_vs_Rho_Eps_phase_diagram}(a), we show the $D$--$\beta$ phase diagram for fixed $\rho_0=5$ and $\epsilon=2.5$. At sufficiently large $\beta$, MIP emerges even for $D=1$, similar to the behavior observed in AIM~\cite{jd2024MIP}, and the MIP region expands further as $D$ decreases. Lower $D$ facilitates the spontaneous formation of transverse-polarity droplets, which act as pinning centers for MIP. It is known that a spontaneous nucleation inside the liquid becomes more likely as $\beta$ increases or $D$ decreases~\cite{benvegnen2023meta,jd2024MIP}. For intermediate $\beta$, the spontaneous nucleation of droplets gives rise to SRO steady states, whereas for small $\beta$, as the separation between hopping and flipping timescales diminishes, the SRO state is no longer observed. Instead, depending on the temperature $(\beta^{-1})$, the system either enters a stripe state or remains in a gaseous state. The stripe state represents a liquid-gas coexistence regime and typically forms when quenching below the liquid spinodal from an ordered state, resulting in several liquid domains. This state predominantly appears at large $\epsilon$, where the strong bias suppresses transverse motion, thereby inhibiting domain coalescence (see lower panel of \texttt{movie09} in Ref.~\cite{zenodo}).

The $D$--$\rho_0$ phase diagram [Fig.~\ref{fig:fig_D_vs_Rho_Eps_phase_diagram}(b)] is shown for $\beta=1$ and $\epsilon=2.5$. This diagram is largely dominated by the short-range order (SRO) phase, with the SRO region expanding and the liquid region diminishing as $D$ is decreased down to $D=0.2$, since lower $D$ enhances the spontaneous emergence of transverse-polarity droplets. At very small diffusion ($D=0.1$), we observe a transition from a gas phase at low densities to a MIP phase at higher densities. A similar occurrence of a MIP-like jammed phase for $D=0.1$ is discussed in Sec.~\ref{sec:droplet}, in the context of Fig.~\ref{fig:fig_rd_D_phasediagram}. This onset of MIP at small $D$ is driven by two competing timescales: a moderately fast flipping process for $\beta=1$ and a slow hopping process associated with $D=0.1$.

The $D$--$\epsilon$ phase diagram [Fig.~\ref{fig:fig_D_vs_Rho_Eps_phase_diagram}(c)] is plotted for $\beta=1$, $\rho_0=5$, and relatively large values of $\epsilon$, as the SRO phase is absent at small $\epsilon$. Although droplet nucleation still occurs at small $\epsilon$, the resulting droplets cannot persist due to enhanced hopping in directions orthogonal to the bias. Moreover, the critical droplet size needed to destabilize the liquid increases with decreasing $\epsilon$ (see Sec.~\ref{sec:droplet}). At $D=1$, increasing $\epsilon$ causes a transition from the liquid phase to a stripe state. Since hopping along the biased directions becomes dominant at high velocities, the transverse extent of the liquid region becomes compressed. The SRO phase emerges in the range $0.2 \leq D \leq 0.7$ across most values of $\epsilon$, except for $D=0.1$, where the MIP state becomes dominant.

Taken together, the phase diagram of the APM~\cite{chatterjee2020flocking,mangeat2020flocking} becomes substantially richer when the diffusion constant $D$ is varied. Beyond the conventional APM phases, we identify two novel states: a short-range ordered (SRO) phase and a motility-induced pinning (MIP) phase, along with a stripe coexistence region not previously reported. Including these with the gas, liquid, and liquid-gas coexistence phases yields a comprehensive understanding of the complex phase behavior in the APM.


\section{Discussion}
\label{conclusion}
Motivated by recent studies highlighting the fragility of the polar ordered phase in various flocking models~\cite{besse2022metastability,benvegnen2023meta,jd2024MIP,karmakar2024consequence,TSAIM}, we have investigated the stability of ordered phases in a discrete-symmetry flocking model, specifically the 4-state APM~\cite{chatterjee2020flocking, mangeat2020flocking}. Our analysis began with an examination of how small counter-propagating and transversely propagating droplets perturb the polar liquid phase. Interactions between these droplets and the background polar flocks generate local disturbances that can result in either a full reversal of the flock or the emergence of a distinct intermediate configuration termed as the {\it sandwich state}, where a reversed domain becomes embedded between two domains of the original flock. The balance between these two outcomes is found to depend critically on the self-propulsion speed $(\epsilon)$, the inverse temperature $(\beta)$, average particle density $(\rho_0)$, and the underlying lattice anisotropy. At high $\beta$, strong spin alignment and reduced transverse fluctuations favor the formation of sandwich states. Conversely, at lower $\beta$, complete reversals become likely within a finite intermediate range of $\epsilon$. An increase in $\epsilon$ reduces transverse diffusion, facilitating the formation of sandwich states by counter-propagating droplets and stabilizing transverse droplets by minimizing leakage across the interface. Additionally, higher $\epsilon$ lowers the threshold droplet size required to initiate either process.

We further confirm that the APM, similar to the AIM~\cite{jd2024MIP,TSAIM}, fails to support long-range polar order over a wide region of its phase space. Under conditions of low diffusion and high directional bias, the APM exhibits spontaneous nucleation of transversely propagating droplets, which fragment a globally ordered flock into multiple smaller domains corresponding to the four internal spin states. At very low diffusion and large $\beta$, this process gives rise to a motility-induced pinning transition, leading to jammed configurations characterized by pinned interfaces, even in the absence of quenched disorder. This phenomenon is consistent with behavior observed in both single-species and multi-species AIM systems~\cite{jd2024MIP,TSAIM}. System size also plays a crucial role in determining metastability. For counter-propagating droplets, increasing the transverse system size suppresses full reversals and favors the emergence of sandwich states. On the other hand, larger systems facilitate faster spontaneous droplet nucleation, particularly for transversely moving small droplets~\cite{benvegnen2023meta}.

Altogether, these results reveal a rich interplay between temperature, diffusion, self-propulsion velocity, and system size in shaping the stability of phases in discrete flocking models such as the APM. We hope that our findings, in conjunction with previous works~\cite{besse2022metastability,benvegnen2023meta,jd2024MIP,karmakar2024consequence,TSAIM}, encourage further exploration into the stability of polar order in a broad range of active matter systems. Beyond theory, our observed phases should be testable in dry active systems where motion is constrained to a few discrete directions by design or confinement. Examples include Quincke rollers guided by square electrode grids, vibrated polar grains on grooved square substrates, NESW-programmed robot swarms, and bidirectional corridor flows (such as ants or pedestrians)~\cite{{geyer2019freezing,bricard2013emergence,lei2023exploring,poissonnier2019experimental,feliciani2016empirical}}.

\begin{acknowledgements}
	MK and RP acknowledge the computational facility provided by the Indian Association for the Cultivation of Science (IACS). SC, MM, and HR are financially supported by the German Research Foundation (DFG) within the Collaborative Research Center SFB 1027. SC and MM acknowledge useful discussions with Jae Dong Noh.
\end{acknowledgements}


\appendix

\section{Temporal evolution of a counter-propagating droplet}
\label{drop_pos}

\begin{figure}[t]
    \centering
    \includegraphics[width=\columnwidth]{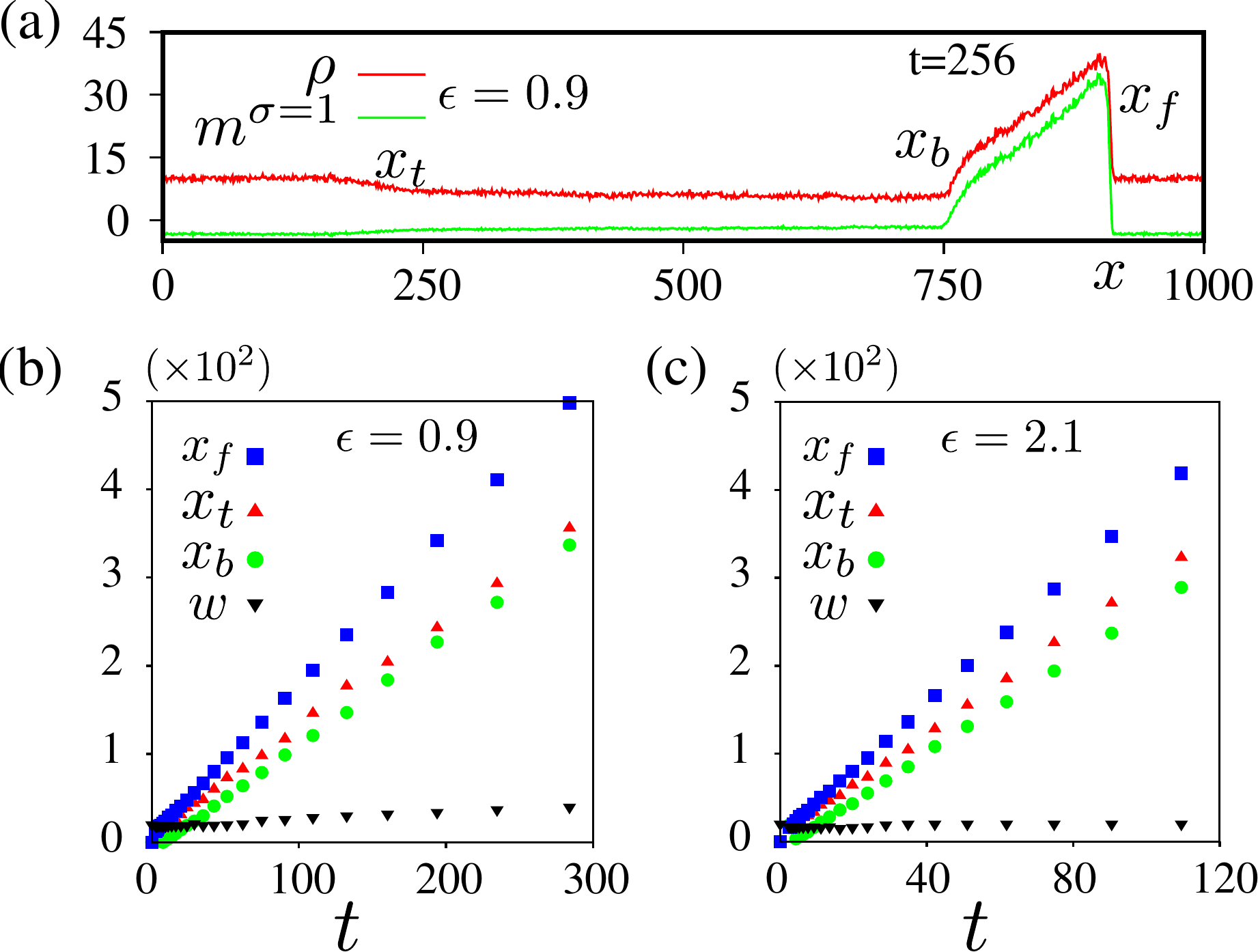}
    \caption{(color online) {\it Temporal evolution of a counter-propagating droplet}. (a)~Density (red line) and magnetization (green line)  profiles for the growing counter-propagating droplet at small self-propulsion velocity ($\epsilon=0.9$). $x_f$, $x_b$, and $x_t$  are the instantaneous positions of the growing droplet front, back, and tail, respectively along the $x$-axis. (b--c)~$x_f$, $x_b$, and $x_t$ are measured relative to the center $(x_c,y_c)$ = $(500,25)$, and the width of the droplet $w$ (measured along the $y$-axis) as a function of time $t$ for small ($\epsilon=0.9$) and large self-propulsion ($\epsilon=2.1$) velocity, respectively. Parameters: $D=1$, $\beta=1$, $\rho_0=10$, $r_d = 10$, $\rho_0^d = 1.2\rho_0$, $L_x=1000$, and $L_y=50$.}
    \label{fig:fig_drop_profile_position}
\end{figure}

\begin{figure*}[tb]
    \centering
    \includegraphics[width=1.9\columnwidth]{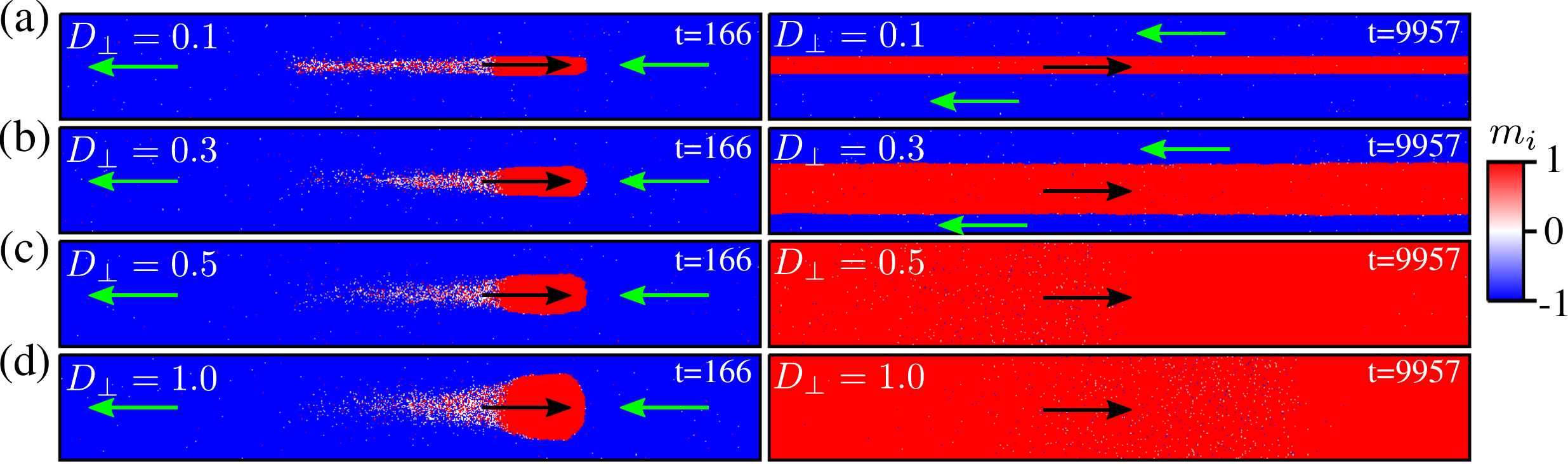}
    \caption{(color online) {\it Formation of the sandwich state in the AIM for small transverse diffusion}. (a) $D_{\perp}=0.1$ and (b) $D_{\perp}=0.3$. Full reversal of the initial liquid phase is observed for larger values of $D_{\perp}$, (c) $D_{\perp}=0.5$, and (d) $D_{\perp}=1$. Colorbar represents per-site magnetization $m_i=n_i^+ - n_i^-$, where $n_i^+$ and $n_i^-$ are the number of particles with $s=+1$ and $s=-1$ at site $i$, respectively. Parameters: $D_\parallel=1$, $\beta=2$, $v=1$, $\rho_0=10$, $r_d=10$, $\rho_0^d=5\rho_0$, $L_x=1000$, and $L_y=100$.}
    \label{fig:fig_aim_drop_nucl}
\end{figure*}

Fig.~\ref{fig:fig_drop_profile_position} illustrates the contrast in the temporal evolution of a counter-propagating droplet at low and high propulsion velocities. In Fig.~\ref{fig:fig_drop_profile_position}(a), the density (red line) and magnetization (green line) profiles of a counter-propagating growing droplet illustrate the spatial distribution of particles and their spin states. Both the density and magnetization profile (droplet state $\sigma=1$) tilt in the front ($x_f$) regions of higher particle concentration due to the droplet nucleation along the propagation. Opposite to the propagation, the height of the profiles is gradually decreased up to $x_b$, and then a low-density disordered region is created up to the tail $x_t$. The temporal evolution of the droplet under small self-propulsion velocity [Fig.~\ref{fig:fig_drop_profile_position} (b)] shows that the positions of the front ($x_f$) and back ($x_b$) interfaces, including the droplet tail ($x_t$) grow rapidly. In addition, the droplet front moves faster than the back and the tail, leading to an overall elongation of the phase constituted by the droplet. Droplet width $w$, which is measured as the difference between the two boundaries of the density profile along the $y$-direction, appears to grow relatively slowly within the ordered phase. For a larger self-propulsion velocity, $\epsilon=2.1$, the droplet moves faster while maintaining a nearly constant droplet width $w$ [Fig.~\ref{fig:fig_drop_profile_position}(c)]. This leads to the emergence of the sandwich state at higher self-propulsion velocities.

\section{Sandwich formation in the active Ising model (AIM) at reduced transverse diffusion}
\label{sandwich_aim}

We simulate the AIM (as defined in Ref.~\cite{benvegnen2023meta}) on a two-dimensional lattice where each particle carries an Ising spin variable, $s = \pm 1$, indicating its self-propulsion direction. We decompose the diffusion constant $D$ into two components, $D_{\perp}$ (along $\pm y$ direction) and $D_{\parallel}$ (along $\pm x$ direction). Particles hop to one of their four neighbors at diffusion rate $2\left(D_{\perp}+D_{\parallel}\right)$, self-propel to a neighboring site on the right $(s=+1)$ or left $(s=-1)$ at rate $v$, and can flip their spin state $(s \to -s)$ at rate $\gamma e^{-\beta s m_i/\rho_i}$, where $m_i=n_i^+ - n_i^-$ and $\rho_i=n_i^+ + n_i^-$ are the local magnetization and density at site $i$, respectively. We set $\gamma=1$ and the Monte Carlo time unit of our simulation is taken as $\Delta t = \left(2D_{\perp}+2D_{\parallel}+v+e^\beta \right)^{-1}$. The artificial droplet excitation follows the protocol described in Sec.~\ref{sec:droplet}.

The initial system is prepared by inserting a droplet of $s=+1$ particles in the high-density liquid phase of $s=-1$ particles. We vary $D_{\perp}$, keeping $D_{\parallel}=v=1$. Fig.~\ref{fig:fig_aim_drop_nucl} shows that a sandwich state can also emerge in the AIM (previously unreported) when $D_{\perp}$ is low.  For $D_{\perp}=0.1$ and $D_{\perp}=0.3$ [Fig.~\ref{fig:fig_aim_drop_nucl}(a--b)], the droplet cannot expand along the transverse $(\pm y)$ direction due to limited diffusion and does not form a comet-like structure, as shown in Ref.~\cite{benvegnen2023meta}, instead, a sandwich state of $s=+1$ (red) and $s=-1$ (blue) particles appears. As $D_{\perp}$ increases [Fig.~\ref{fig:fig_aim_drop_nucl}(c--d)], the comet-like structure of the droplet reappears due to enhanced transverse diffusion, resulting in a complete reversal of the initial liquid phase, as reported in Ref.~\cite{benvegnen2023meta}. 

As demonstrated in Fig.~\ref{fig:fig_aim_drop_nucl}, the outcome of an artificial droplet excitation, whether it leads to a sandwich state or a full reversal of the initial liquid phase, depends on the transverse diffusion $D_{\perp}$, where the biased movement of the droplet is confined in $1d$ ($\pm x$ direction). The transverse length of the system $L_y$ also plays a crucial role, as a complete phase reversal requires the growing droplet front to connect the boundaries along the transverse direction ($y=1$ and $y=L_y$), which is only possible if sufficient transverse diffusion is present. For $D_{\perp}=1$ (maximum value), the droplet front will inevitably connect the transverse boundaries and trigger a full reversal, regardless of $L_y$. However, when $D_{\perp}$ is small, a full reversal can only occur if $L_y$ is also small. Therefore, the critical $D_{\perp}$, above which a full reversal takes place, increases with $L_y$. For instance, in Fig.~\ref{fig:fig_aim_drop_nucl}(b), if $L_y=40$ is considered instead of $L_y=100$, a full reversal is observed rather than the sandwich state. However, if $L_y=40$ is maintained and $D_{\perp}$ is reduced, the sandwich state reappears.

\section{Impact of spatial anisotropy on the emergence of the APM sandwich state}
\label{sandwich}

\begin{figure}[t]
    \centering
    \includegraphics[width=\columnwidth]{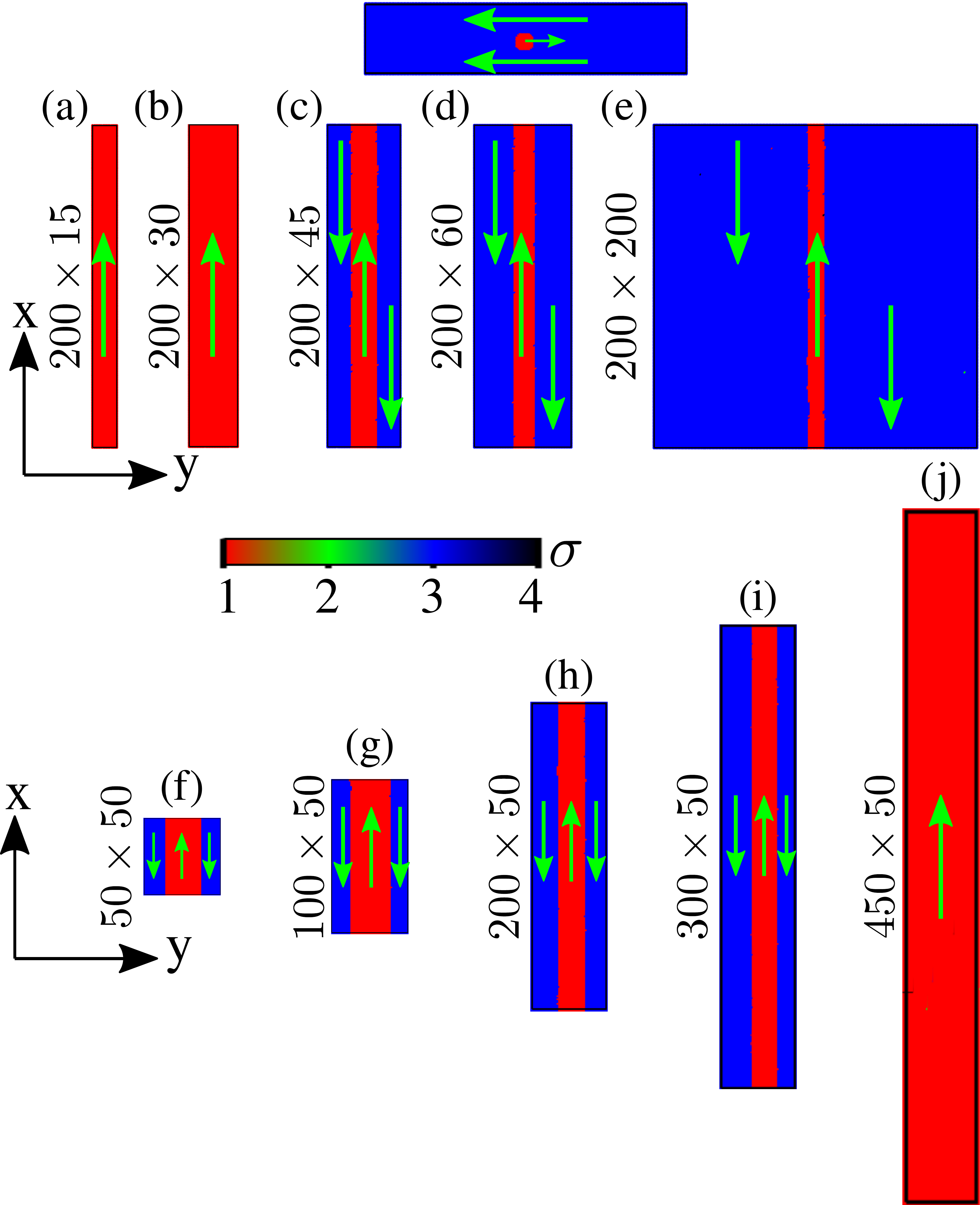}
    \caption{(color online) {\it Sandwich state with spatial anisotropy.} Snapshots exhibiting (a-e) the transition from a fully reversed phase to a sandwich state as $L_y$ increases $(L_x=200)$ and (f--j) the transition from a sandwich state to a fully reversed phase as $L_x$ increases $(L_y=50)$ with counter-propagating droplet excitations. A typical initial state is shown at the top of the figure. Parameters: $D=1$, $\beta=1$, $\epsilon=0.9$, $\rho_0=10$, $r_d = 6$, and $\rho_0^d = 5\rho_0$.}
    \label{fig:fig_dropsandwich_Ly}
\end{figure}

Fig.~\ref{fig:fig_dropsandwich_Ly} shows a series of steady-state snapshots after introducing a counter-propagating droplet of state $\sigma=1$ (red) into a polar liquid phase of $\sigma=3$ state particles. Fig.~\ref{fig:fig_dropsandwich_Ly}(a--e) demonstrates the transformation of the steady-state from a full reversal of the initial polar-ordered liquid phase to a sandwich state as the system size along the transverse direction $(L_y)$ increases for fixed $L_x=200$. For $L_y=15$ and $L_y=20$, the system undergoes a complete reversal of its old liquid phase of $\sigma=3$ to a new uniform ordered phase of $\sigma=1$ [Fig.~\ref{fig:fig_dropsandwich_Ly}(a--b)]. In subsequent snapshots at larger transverse system sizes $L_y=45$, $L_y=60$, and $L_y=200$ [Fig.~\ref{fig:fig_dropsandwich_Ly}(c--e)], the excitation of the opposite polarity droplet can not fully reverse the original ordered phase and the system stabilizes into a sandwich state of $\sigma=3$ (state of the initial liquid) and $\sigma=1$ (droplet state) state regions. However, Fig.~\ref{fig:fig_dropsandwich_Ly}(f--j) exhibits the reverse scenario when $L_x$ is increased, keeping $L_y$ fixed where we observe the steady-state transforms from a sandwich state [Fig.~\ref{fig:fig_dropsandwich_Ly}(f--i)] to a full reversal phase [Fig.~\ref{fig:fig_dropsandwich_Ly}(j)] with $L_x$. This highlights the profound impact of spatial anisotropies on the stability of the polar ordered phase in the 4-state APM.

\section{Profiles and stability of the sandwich state}
\label{sw_stability}

\begin{figure}[t]
    \centering
    \includegraphics[width=\columnwidth]{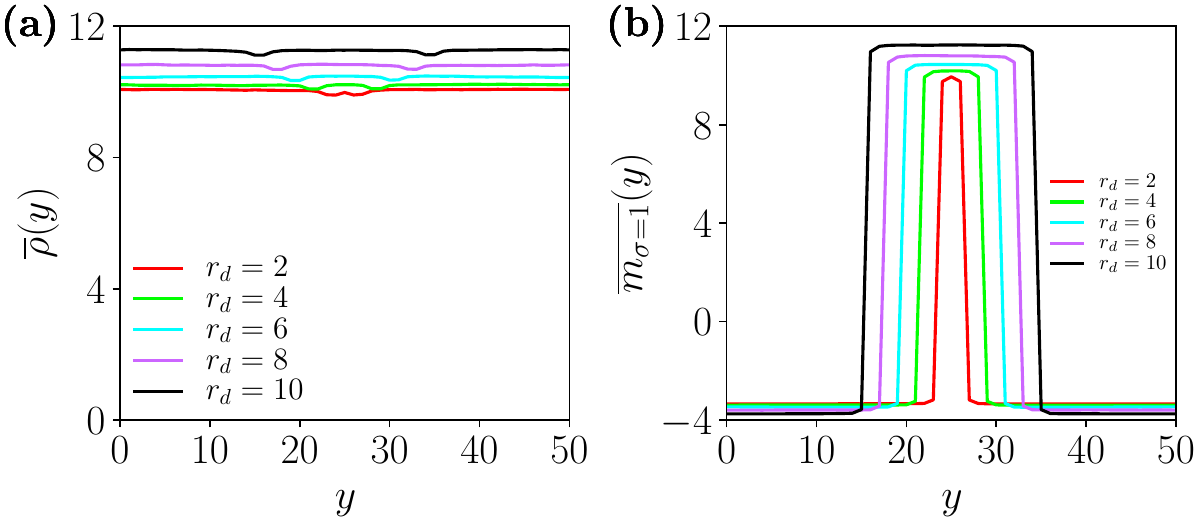}
    \caption{(color online) {\it Density and magnetization profiles of the sandwich state.} Time-averaged (a) density and (b) magnetization profiles of the sandwich state for increasing droplet radius $r_d$. Parameters: $D=1$, $\beta=1$, $\epsilon=2.7$, $\rho_0=10$, and $\rho_0^d = 5\rho_0$. $L_x=200$, $L_y=50$.}
    \label{fig:fig_sandwich_profile}
\end{figure}

Fig.~\ref{fig:fig_sandwich_profile} shows the time-averaged density and magnetization profiles, integrated over the x-coordinate, of the sandwich state formed by droplets of $\sigma=1$ embedded in a $\sigma=3$ liquid, for various droplet radii $r_d$. For all $r_d$, the density $\overline{\rho}(y)$ is almost uniform (except near the interfaces) around the average density after introducing the droplet: $\rho_{\rm tot} = (N + \Delta N)/L_xL_y$, whereas the magnetization $\overline{m_{\sigma=1}}(y)$ peaks at the middle, where the droplet have been introduced, and becomes negative outside where the magnetization $\overline{m_{\sigma=3}}(y)$ dominates. Therefore, the sandwich state is not density-segregated, and only the internal polarization varies along the transverse direction.

\begin{figure}[t]
    \centering
    \includegraphics[width=\columnwidth]{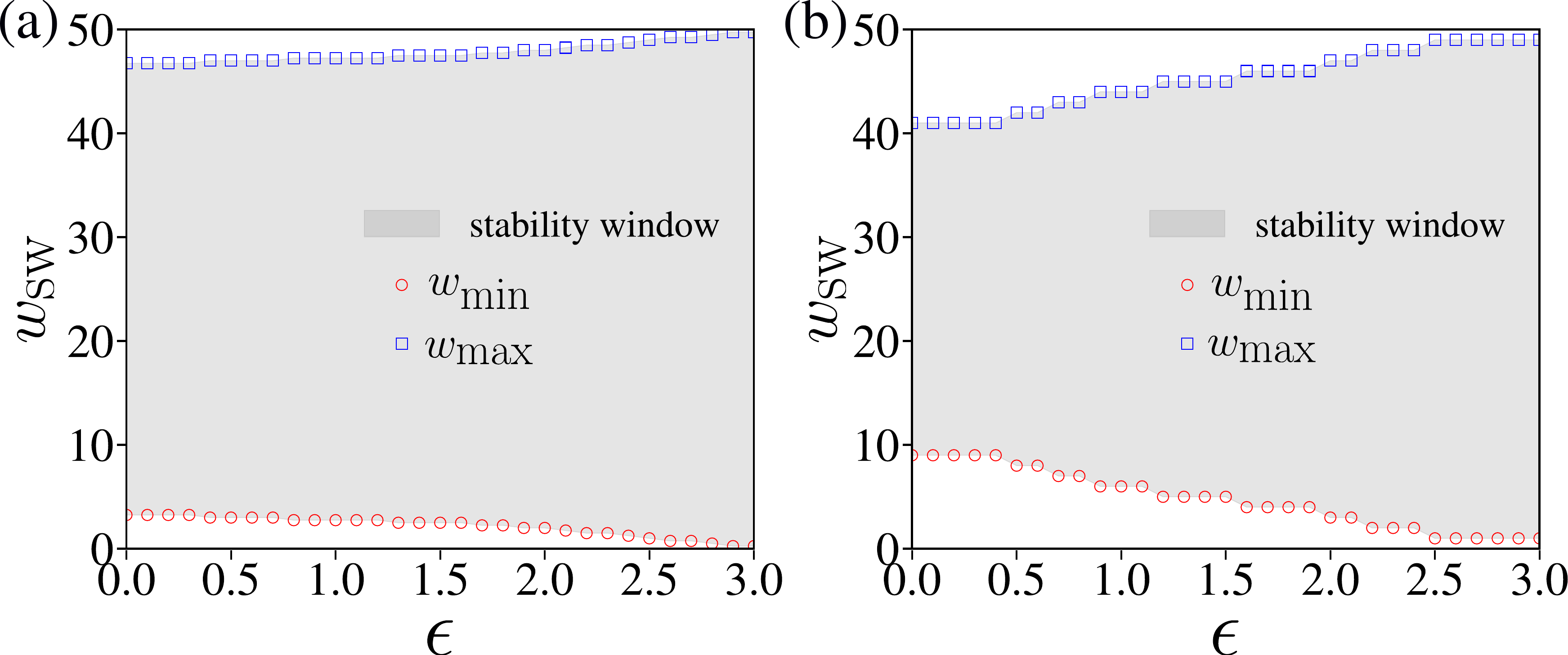}
    \caption{(color online) {\it Stability of the sandwich state.} (a)~Hydrodynamic theory for $D=1$, $\beta=0.9$, $\rho_0=3$, $L_x=200$, and $L_y=50$. (b)~Numerical simulations for $D=1$, $\beta=1$, $\rho_0=10$, $L_x=200$, and $L_y=50$. At fixed $\epsilon$, the sandwich state is stable if the width $w_{\rm sw}$ is between $w_{\min}$ and $w_{\max}$. For a prepared sandwich, widths below $w_{\min}$ collapse to a uniform single-polarity liquid matching the outer domains, while widths above $w_{\max}$ expand and convert the entire system to a liquid phase of inner polarity.}
    \label{fig:fig_sandwich_stability}
\end{figure}

Fig.~\ref{fig:fig_sandwich_stability} summarizes the stability of the sandwich state by solving the hydrodynamic equations, Eq.~\eqref{PDEhydro0} [Fig.~\ref{fig:fig_sandwich_stability}(a)]; and from numerical simulations [Fig.~\ref{fig:fig_sandwich_stability}(b)]. For fixed $\epsilon$, we analyze the stability a sandwich state with initial width $w_{\rm sw}$, where the rectangular domain is separated into two vertically-organized lanes of state $\sigma=1$ and $\sigma=3$ of size $w_{\rm sw}$ and $L_y-w_{\rm sw}$, respectively, with equal density. The sandwich state remains stable only if the initial width is between two thresholds: a minimal width $w_{\min}(\epsilon)$ and a maximal width $w_{\max}(\epsilon)$. If $w_{\rm sw} < w_{\min}$, the system goes to a liquid of state $\sigma=3$, and if $w_{\rm sw} > w_{\max}$, the system goes to a liquid of state $\sigma=1$. With symmetry arguments, we can identify the relation $w_{\max}=L_y-w_{\min}$. The hydrodynamic and numerical results agree qualitatively on this stability analysis.

\section{Counter-propagating droplet dynamics for varying $\beta$}
\label{CD_beta}

\begin{figure}[t]
    \centering
    \includegraphics[width=\columnwidth]{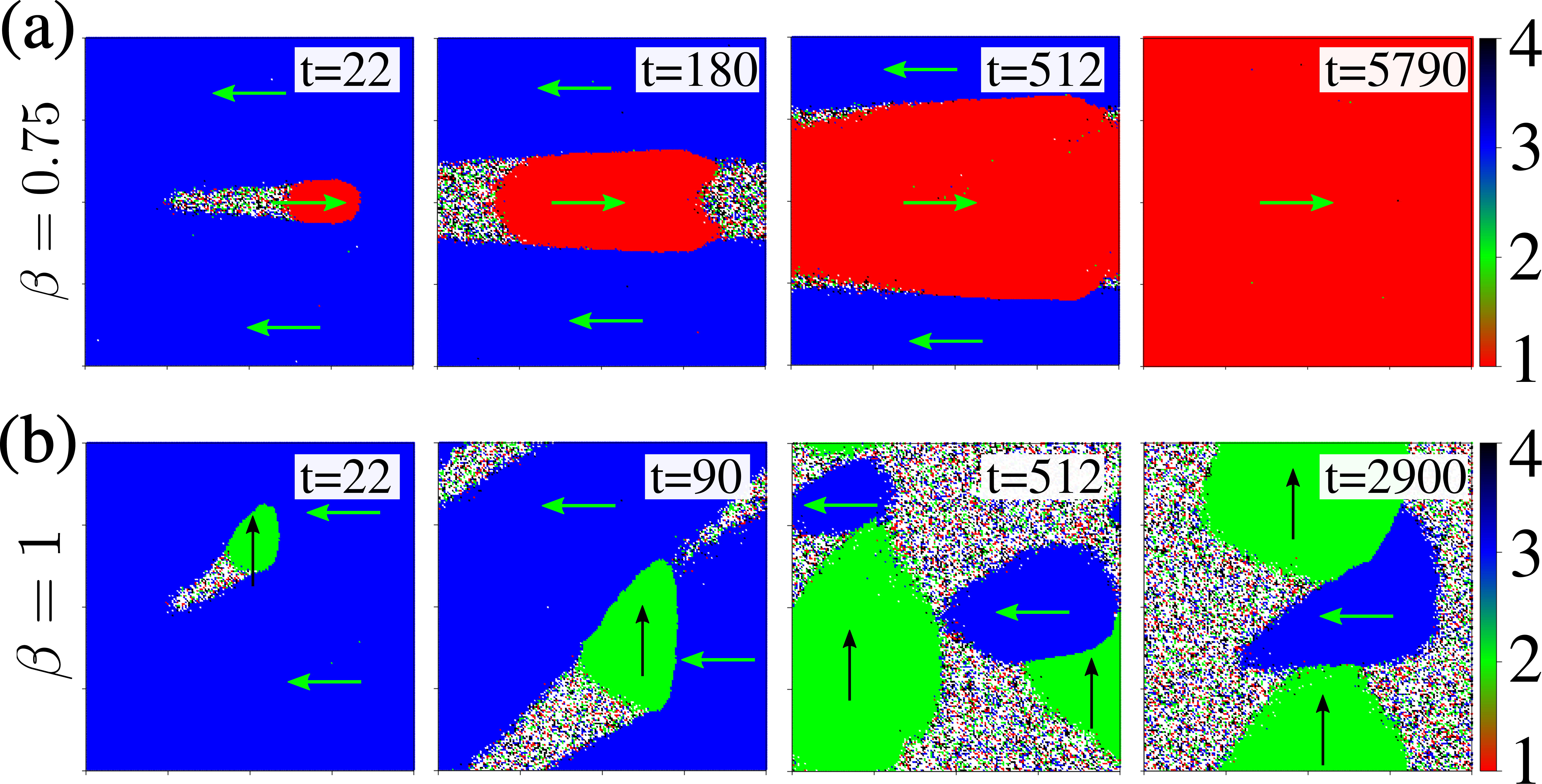}
    \caption{(color online) {\it Time evolution for droplet excitations for intermediate particle velocity}. (a) Counter-propagating droplet for $\beta = 0.75$. Similar to Fig.~\ref{fig:apm_ftcs_snaps}(b), the initial ordered phase $(\sigma = 1)$ is fully replaced by the droplet-induced phase $(\sigma = 3)$. (b) Transversely-propagating droplet for $\beta=1$. Similar to Fig.~\ref{fig:apm_ftcs_snaps}(d), the system evolves into a steady-state defined by two high-density clusters, one from the initial liquid (blue, $\sigma=3$) and the other from the droplet (green, $\sigma=2$), that move in mutually orthogonal directions. Parameters: $D=1$, $\epsilon=1.5$, $\rho_0 = 10$, $r_d = 10$, $\rho_0^d = 5\rho_0$, $L_x = L_y = 200$. Arrows indicate the direction of motion, and the colorbar represents $\sigma$.}
    \label{fig:cd_beta}
\end{figure}

In Sec.~\ref{sec:droplet}, we examined the effect of artificial droplet excitation deep within the ordered phase of the APM by considering large $\beta = 1$ and density $\rho_0 = 10$. We found that, under these conditions, counter-propagating droplet excitation leads to a steady-state characterized by a partially reversed sandwich configuration, in the absence of any lattice anisotropy. In Fig.~\ref{fig:cd_beta}(a), we explore how this behavior changes when $\beta$ is reduced, thereby enhancing thermal fluctuations in a square geometry ($L_x=L_y=200$). For $\beta>0.8$, we find that the system consistently exhibits the sandwich state across all values of $\epsilon$, with the droplet-induced lane becoming narrower as $\epsilon$ increases. For $\beta=0.75$ and $\epsilon=1.5$ [Fig.~\ref{fig:cd_beta}(a)], the counter-propagating droplet is observed to completely replace the initial ordered phase. This full reversal occurs within a central range of $\epsilon$ values $(1.2 \leq \epsilon \leq 1.8)$, near the reorientation transition~\cite{chatterjee2020flocking,mangeat2020flocking}, which itself depends on $\beta$. Outside this range, the system predominantly exhibits a sandwich state. Fig.~\ref{fig:cd_beta}(b) shows the time-evolution of the system with transverse droplet excitation at $\beta=1$ and $\epsilon=1.5$ where the steady-state is characterized by mutually orthogonal motion of two high-density clusters of the droplet (green) and the initial liquid state (blue). This mutual orthogonality in motion, reinforced by strong directional bias and alignment interactions, allows the two high-density clusters to stably coexist without coalescing, even over long times.

\section{Metastability in the APM with a different hopping mechanism}
\label{newAPM}

\begin{figure}[t]
    \centering
    \includegraphics[width=\columnwidth]{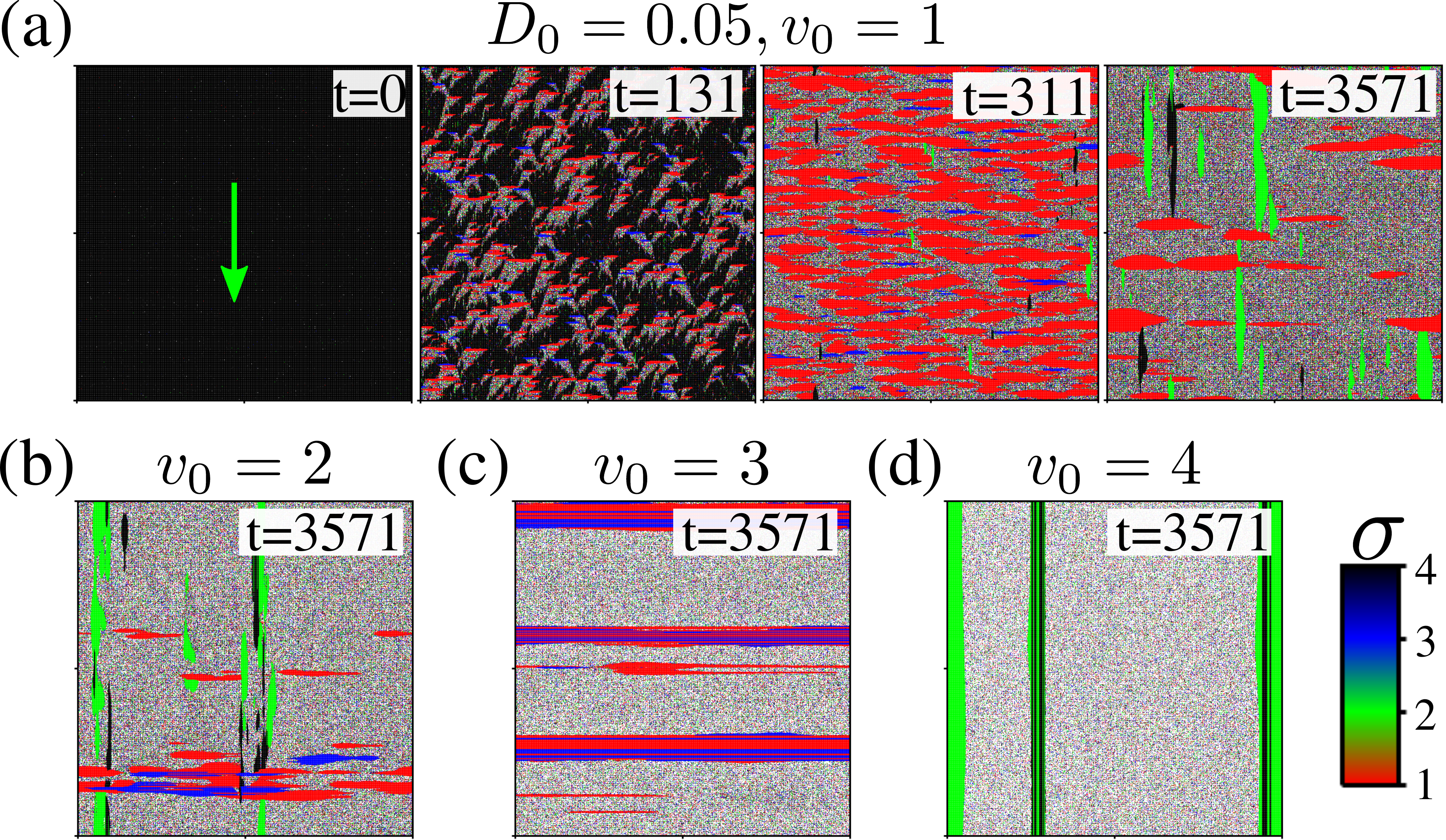}
    \caption{(color online) {\it Spontaneous nucleation in the APM with a different hopping mechanism.} (a) Time evolution snapshots of the modified APM (see Appendix \ref{newAPM}) with initially ordered configuration, $\sigma=4$  (black domain), exhibit spontaneous nucleation of droplets which indicate fragility of the LRO phase similar to Fig.~\ref{fig:fig_spontaneous_nucleation}. (b--d) Steady-state snapshots with increasing $v_0$ show reduced spontaneous nucleation of droplets and formation of longitudinally moving bands (signifying liquid-gas coexistence region), which characterizes high-velocity particle regimes. Parameters: $D_0=0.05$, $\beta=1$, $\rho_0=5$, and $L_x = L_y = 1024$. The four different colors in the color bar represent the different spin states.}
    \label{fig:fig_manual_nucleation}
\end{figure}

Inspired by the AIM model proposed in Ref.~\cite{jd2024MIP}, here we propose a moderately different version of the APM and show that, similar to the AIM, this model also exhibits spontaneous nucleation of droplets and metastability of the LRO phase. In this model, we change the hopping mechanism by keeping the flipping rate unchanged [Eq.~\eqref{flipeq}]. The particles now hop to one of their four neighbors at a rate of $4D_0$ and self-propel to a neighboring site according to spin state $\sigma$ at a rate $v_0$. Therefore, the total hopping rate becomes $4D_0+v_0$ ($4D$ in the previous model). This hopping is equivalent to Eq.~\eqref{whop} with $D_0=D(1-\epsilon/3)$ and $v_0=4D\epsilon/3$. Since $\epsilon$ is constrained within the range $0 \leqslant \epsilon \leqslant 3$, the velocity $v_0=4D_0\epsilon/(3-\epsilon)$ in this model is theoretically unbounded (at $\epsilon=3$, $v_0 \to \infty$). The Monte Carlo time unit is also adjusted and is now given by $\Delta t = \left[4D_0 + v_0 + \exp(4\beta)\right]^{-1}$ with $4D_0 + v_0=4D$.

\begin{figure}[t]
    \centering
    \includegraphics[width=\columnwidth]{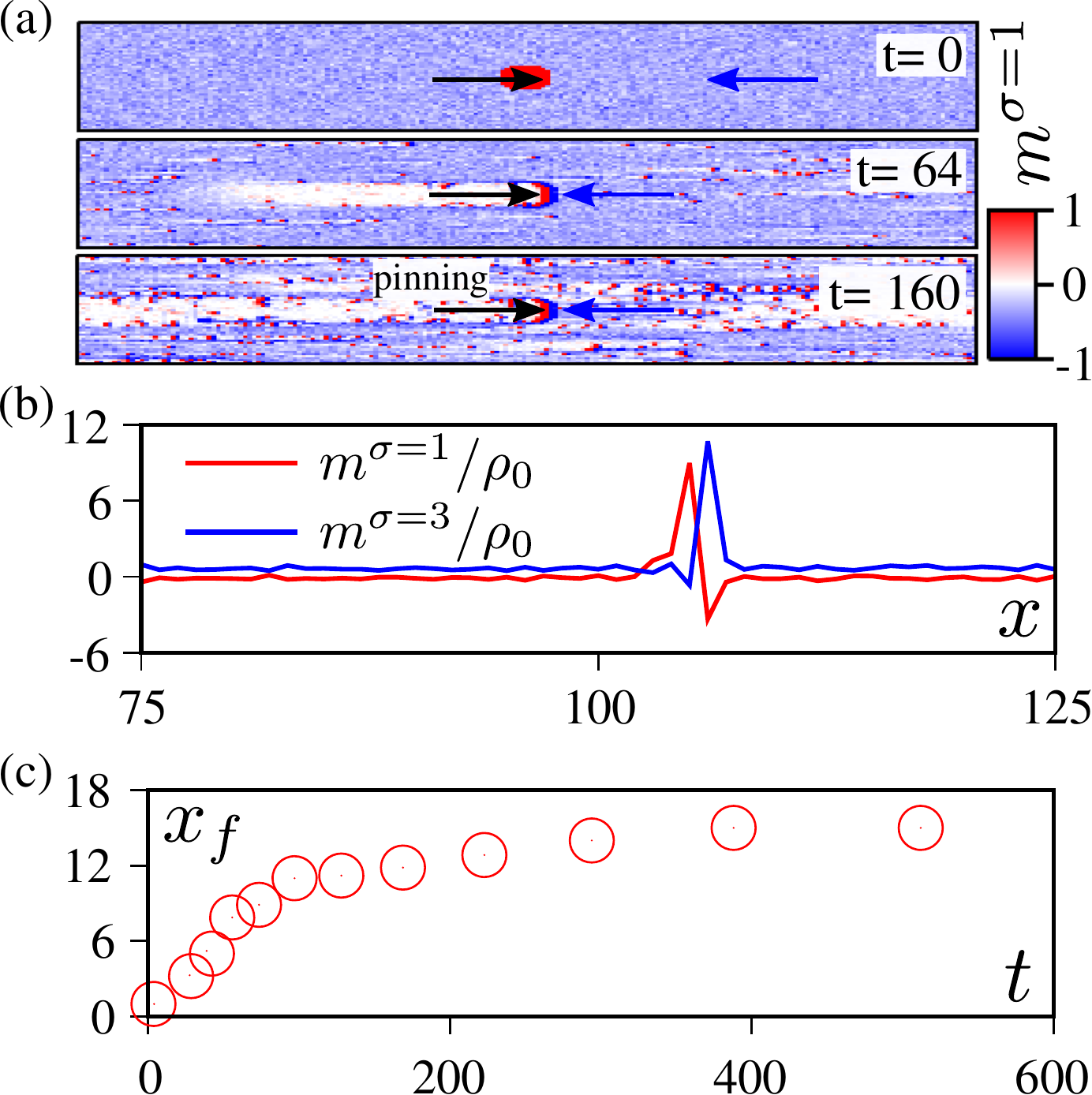}
    \caption{(color online) {\it Droplet jamming at small temperatures.} (a) Snapshots at different times, $t$, showing droplet pinning in the 4-state APM. A droplet of $\sigma=1$ particles, initially placed in an ordered region of $\sigma=3$ particles, becomes pinned over time, highlighting the pinning effect under low diffusion and high $\beta$. A movie (\texttt{movie11}) of the same can be found at Ref.~\cite{zenodo}. (b) Magnetization profile at the interface between the droplet state (red solid line) and the ordered phase (blue solid line) at $t=160$, showing a pronounced peak that indicates a jammed, high-magnetization region at the droplet front. (c) Position of the droplet interface front ($x_f$) relative to the center, showing saturation over time, indicating the formation of a jammed, pinned interface. Parameters: $D=0.3$, $\beta = 2$, $\epsilon=2.5$, $\rho_0 = 10$, $r_d = 6$, $\rho_0^d = 5\rho_0$, $L_x = 200$, $L_y = 50$.}
    \label{fig:fig_drop_mip}
\end{figure}

Fig.~\ref{fig:fig_manual_nucleation} shows the time-evolution [Fig.~\ref{fig:fig_manual_nucleation}(a)] and steady-state snapshots [Fig.~\ref{fig:fig_manual_nucleation}(b--d)] for various values of self-propulsion velocity $v_0$. Fig.~\ref{fig:fig_manual_nucleation}(a) shows that the high-density liquid phase is also not stable under the new hopping mechanism with diffusion $D_0=0.05$ and $v_0=1$ $(\epsilon=2.5)$ akin to the AIM~\cite{jd2024MIP}. The liquid band, composed of particles in the $\sigma=4$ state (black domain), initially moves downward. Subsequently, spontaneous nucleation of clusters predominantly composed of $\sigma=1$ (red) and $\sigma=3$ (blue) state particles, moving right and left, respectively, emerges within high-density regions ($t=131$). This nucleation disrupts the long-range order (LRO), leading to a new steady state characterized by short-range ordered (SRO) clusters of particles from all states ($t=311$ and $t=3571$). Due to the high velocity, the clusters move longitudinally without significant lateral diffusion. This behavior persists at $v_0=2$ $(\epsilon \simeq 2.73)$ [Fig.~\ref{fig:fig_manual_nucleation}(b)], but for $v_0=3$ $(\epsilon \simeq 2.81)$ [Fig.~\ref{fig:fig_manual_nucleation}(c)] and $v_0=4$ $(\epsilon \simeq 2.86)$[Fig.~\ref{fig:fig_manual_nucleation}(d)], steady longitudinal bands form. This is because increasing $v_0$ increases the total hopping rate, which in the original APM depended only on $D$ and was independent of $\epsilon$ [see Eq.~\eqref{whop}]. A liquid phase normally arises at large $\beta$, because the state flipping rate is higher, and for spontaneous nucleation to occur, the total hopping rate must be small relative to $W_{\rm flip}$, which depends solely on $\beta$ and local density. As $v_0$ increases, both the biased and total hopping rates increase, stabilizing the band motion characteristic of the coexistence region. Therefore, it can be concluded that as the total hopping rate becomes comparable to the flipping rate, spontaneous nucleation ceases to exist. This explains why spontaneous nucleation was not observed for $D=1$ in the original APM~\cite{chatterjee2020flocking,mangeat2020flocking}, since varying $\epsilon$ does not affect the total hopping rate, which remains fixed at $4D$. Spontaneous nucleation in the APM with the original hopping rule can only be observed by reducing the diffusion constant $D$, as shown in Fig.~\ref{fig:fig_spontaneous_nucleation}.

\section{Droplet pinning at low temperature}
\label{dropletBETA}

Fig.~\ref{fig:fig_drop_mip} demonstrates the fate of an inserted droplet within the polar-ordered phase at large $\beta$ and small $D$. Over time, the droplet freezes [Fig.~\ref{fig:fig_drop_mip}(a)] with its front becoming jammed due to the accumulation of opposite-polarity particles $(t=64$ and 160), caused by the motility-induced interface pinning. This phenomenon is further explained by the magnetization profiles of the ordered state ($\sigma=3$) and the droplet state ($\sigma=1$) shown in Fig.~\ref{fig:fig_drop_mip}(b), where the sharp opposite peaks of the two magnetization profiles indicate a high-density interface created by opposite polarity particles. Fig.~\ref{fig:fig_drop_mip}(c) shows the position of the droplet front ($x_f$) over time, exhibiting that after an initial movement in the biased direction, the droplet front remains stationary. A similar pinning of a droplet has been observed in the active $p$-state clock model for $p \leq 4$~\cite{jd2024MIP}. However, in the $p=8$ state, where particles have more orientational degrees of freedom, droplet pinning does not occur~\cite{jd2024MIP}.

\bibliography{ref_meta}

\end{document}